\documentclass{emulateapj}
\usepackage[dvips]{rotating}
\begin{document}
\input epsf.tex

\title{Star Formation Histories of Nearby Elliptical Galaxies. I. Volume
Limited Sample}
\author{Justin H.\ Howell}
\affil{UCO/Lick Observatory, Department of Astronomy \& Astrophysics,\\
University of California, Santa Cruz, California 95064, USA\\
Email: {\tt jhhowell@ucolick.org}
\footnote{now at the Infrared Processing and
Analysis Center, Mail Stop 100-22, California Institute of Technology, Jet
Propulsion Laboratory, Pasadena, CA 91125; jhhowell@ipac.caltech.edu}}

\begin{abstract}

This work presents high $S/N$ spectroscopic observations of a representative
sample of nearby elliptical galaxies.  These
observations provide a strong test of models for the formation
of elliptical galaxies and their star formation histories.  
Combining these data with the Gonzalez (1993)
data set, a volume limited sample of 45~galaxies has been defined.
Results are in agreement with previous studies: the existence of the
metallicity hyper-plane and the Z-plane of Trager et~al. (2000) is
confirmed, and the distribution is clearly due to physical
variations in stellar population parameters and not measurement uncertainty.
Trends between stellar population parameters and galaxy structural
parameters suggest that angular momentum may determine the chemical
abundance of a galaxy at a given mass.

\end{abstract}


\section{Introduction}

Once thought to be simple old and metal-rich systems \citep{ch1baade},
elliptical galaxies have been studied extensively over the past thirty
years.  A number of studies (e.~g. Gonzalez 1993, hereafter G93;
Rose et~al. 1994; J{\o}rgensen 1997; Tantalo, Chiosi, \& Bressan 1998;
Kuntschner \& Davies 1998; Trager et~al. 2000a; Longhetti et al. 2000;
Terlevich \& Forbes 2002; Caldwell, Rose, \& Concannon 2003) 
have shown that elliptical galaxies span a wide range of ages, with a 
relatively small spread in metallicity [Z/H].  The enhancement ratio,
$[\alpha$/Fe], of Mg and nucleosynthetically related elements relative to 
iron peak elements is also found to be significantly greater than solar.

\citet[][hereafter TFWG]{ch1tfwg2} found that elliptical galaxies occupy a
two-dimensional ``metallicity hyper-plane" in the space of [Z/H], 
$[\alpha$/Fe], log~$t$, and log~$\sigma$, where $t$ is the light-weighted 
single stellar population (SSP) equivalent age and $\sigma$ is the velocity 
dispersion.  Although cluster galaxies are older on average than field 
galaxies, a wide range of ages is seen in both environments \citep{jorg99}.  
\citet{geb03} suggested that a frosting of continuing star formation is 
necessary to explain the lack of color evolution in
early type field galaxies since $z = 1$.  TFWG also proposed a frosting
model of star formation, as will be discussed in more detail in \S~3.4. This 
result emphasizes the important
point that the analysis of galaxy properties using SSP models does not
accurately reflect the actual star formation histories of most galaxies.
As TFWG and \citet{geb03} also note, the observables do not discriminate well
between different star formation histories.  In multiple stellar population
models, different combinations of mass fractions, ages, and metallicities can
produce observationally indistinguishable galaxies.

Because the $\alpha$ elements are primarily produced in SN II
while SN Ia produce mostly Fe, $[\alpha$/Fe] may reflect the duration of star 
formation assuming a constant, universal IMF.  In high-redshift starburst, 
[$\alpha$/Fe] will be high, as no SN Ia could have gone off to introduce 
additional Fe.  In contrast, the long and continuous star formation histories 
typical of spiral galaxies produce $[\alpha/{\rm Fe}]\sim0$ as a result of 
plentiful SN Ia enrichment over billions of years.  Thus abundance ratios 
might constrain the formation mechanism of individual galaxies.  Two extreme 
formation mechanisms are a fast clumpy collapse model and a spiral merger
model.  The fast clumpy collapse model is in essence a refinement of the
classical \citet{ch1larson} monolithic collapse scenario: protogalactic gas
clouds collide and merge at high redshift, forming stars rapidly as the gas
dissipatively collapses to the center of the system.  In a major merger
of two spirals, the bulk of the total mass is already in stars formed
over long timescales, while the remaining gas is metal-rich with near solar
[$\alpha$/Fe].  An elliptical formed in a fast clumpy collapse
would have high global [$\alpha$/Fe] values
since most stars formed from gas enriched by only a handful of SN II.

Previous spectroscopic data sets \citep[e.~g., G93,][]{ch1kd98}
do not necessarily include a representative sample of elliptical galaxies
in the local universe.  Thus conclusions based on such data sets may
reflect some unanticipated bias in sample selection rather than physical
relations between structural and/or stellar population parameters of
elliptical galaxies.  Stellar population modeling has advanced considerably
in recent years: the inclusion of nonsolar abundance ratios among 
$\alpha$-elements as pioneered by \citet{ch1tb95} and \citet{ch1tfwg1} has been 
included in more recent models such as \citet*[][hereafter TMB]{ch1tmb03}.

This paper is organized as follows.  Sample selection, observations, and 
data reduction are described in \S~2.  The data are analyzed and stellar 
population parameters are derived in \S~3.  Conclusions are presented in \S~4.

\section{The Data}
\subsection{Sample Selection}

A volume--limited sample is defined, consisting of all 45 early type 
(E or S0) galaxies with $m-M<32.5$ ($v<2200$ km ${\rm s}^{-1}$
for $H_0=70$ km ${\rm s}^{-1}$ ${\rm Mpc}^{-1}$), $M_B<-19.5$,
$\delta>-20^{\circ}$, and $|b|>15^{\circ}$.  The galaxies were
selected using the NASA/IPAC Extragalactic Database (NED).
The cumulative luminosity distribution of this sample is consistent 
with that of a Schechter function \citep{schechter} with
the standard parameters of $M^{\ast}=-20.5$ and $\alpha=-1.07$.  This
suggests that the sample proposed here provides a reasonable sampling
of nearby elliptical galaxies within a magnitude of $M^{\ast}$, and that
the sample does not suffer from significant incompleteness near the faint
limit.  The distance measurements used in selecting this sample are
from \citet{ch1tonry01}.  The volume--limited sample is presented in 
Table~\ref{vlsample}.  Many galaxies were previously observed by 
G93, and are noted as such.  For newly observed galaxies we list $S/N$ per 
pixel near the H$\beta$ line.  Although the sample has
somewhat lower $S/N$ than the G93 sample, this has only a small effect on
the total observational errors due to the typical size of calibration
uncertainties; see \S2.6.  The 10 galaxies that were not observed are 
consistent with being drawn from the same population as the 35 galaxies that
were observed, as judged by one-dimensional Kolmogorov-Smirnov (K-S) tests on 
their distribution in velocity dispersion and $M_B$ magnitude.

\subsection{Observations}

The observing program used the KAST spectrograph with the 1200 lines/mm
grating blazed at $5000\mbox{\AA}$ on the Lick 3~meter
telescope.  Observations consist of $4-6$ 25~minute exposures on a
galaxy, usually with the $145''$-long slit oriented along the galaxy's major 
axis, interspersed with 5~minute ``blank'' sky exposures on fields several 
arcminutes away from the target galaxy.
The spectral range $4200\mbox{\AA}$--$5600\mbox{\AA}$ was observed,
with instrumental resolution of approximately 100~km/s.  The velocity 
dispersions of the galaxies are typically $\sim200$~km/s.  Early observing
runs suffered from slightly poorer resolution as a result of suboptimal
spectrograph focus.  The slit width was typically $1.5^{\prime\prime}$,
though in poor seeing conditions this was increased to $2^{\prime\prime}$.
Spectral resolution was measured independently for each observing run
to account for changes in resolution due to different slit widths or focus
settings.  The plate scale for the spectrograph in this configuration was 
$1.17\mbox{\AA}$/pixel in the dispersion direction and 
$0.8^{\prime\prime}$/pixel in the spatial direction.

Numerous Lick index standard star observations were made, drawing from the 
set of stars in common between the Lick/IDS standards of \citet{ch1worthey94}
and the \citet{j99} standards.

\subsection{Data Reduction}

Flat fields were created for each night of each observing run.
The dome flat input images were averaged in the spatial
direction to one dimension, and this median dispersion axis was then 
expanded back to the full size of the CCD.  The final flat field images 
used in this work are the input flat fields divided by the one 
dimensional median for each row of the CCD.  These flat field images
correct for pixel-to-pixel quantum efficiency variations only.  The
data do not extend to red wavelengths where CCD fringing corrections
become important.

Proper sky subtraction is crucial for accurately measuring line strengths
in galaxies.  Oversubtraction of sky flux creates spuriously large line
strengths, while undersubtraction has the reverse effect.  
For all but the most extended galaxies, the ends of the slit provide 
adequate sky subtraction.  This simple technique was used where applicable
for the galaxies observed during or after 2002.  The most extended galaxies,
and those observed prior to 2002, were reduced using the offset sky 
spectra interspersed with the galaxy observations.  The observations
were not taken under photometric conditions, and the sky was found to
vary unpredictably over timescales much shorter than that of a typical
galaxy exposure.  As a result, averaging the sky spectra taken immediately
before and after each galaxy spectrum and scaling the sky spectrum by the
ratio of exposure times is not sufficient for accurate sky subtraction.
The proper amount of sky continuum was estimated by forcing the sky-subtracted
radial brightness profile to conform as closely to an $r^{1/4}$ law 
\citep{dv48} as possible, as explained below.  \citet{burkert} showed
that the $r^{1/4}$ law provides an excellent fit to all elliptical galaxies
between $0.1\leq{r/r_e}\leq 1.5$.

First, the IRAF\footnote{IRAF is distributed by the National Optical Astronomy
Observatories, which are operated by the Association of Universities for
Research in Astronomy, Inc., under cooperative agreement with the
National Science Foundation.}
task {\sc geomap} was used to rectify the galaxy spectrum
such that a fixed spatial location along the slit lies at a constant $y$~pixel
coordinate.
Cosmic rays and bad pixels were removed at this step.  
The median of the galaxy spectrum images was subtracted from each
individual frame.  The resulting difference frames were then divided 
by the median image plus a floor value equal to ten times the standard 
deviation at the edges of the slit.  The resulting fractional deviation 
image was used to flag deviant pixels by interactively setting a 
threshold value.  The flagged pixels and all neighboring pixels were 
masked from the input images, which
were then combined to form the galaxy spectrum.  
Each input exposure was corrected for frame to frame variations in the 
background flux, putting every exposure on a common relative sky continuum
scale.  Specifically, the combined galaxy spectrum was subtracted from each
cosmic ray corrected input spectrum.  The statistics of the difference image
near each edge of the slit were used to estimate a relative sky offset
between the input images.  The measured offsets were added or subtracted
from each spectrum to place them all on a common relative sky continuum 
scale prior to measuring the absolute sky subtraction amount needed to
fit an $r^{1/4}$ profile.  Since sky intensity varies slowly with 
wavelength, a wavelength-independent offset was used in order to best 
match the sky continuum levels near the H$\beta$, Mg$b$, Fe5270, and Fe5335 
lines.
Next, the combined galaxy spectrum was collapsed along the dispersion axis 
in two wavelength ranges, one around the wavelength of H$\beta$ and
one around the wavelengths of the Mg$b$, Fe5270, and Fe5335 lines, to 
create radial intensity profiles.
An $r^{1/4}$ profile was fit to the galaxy radial profiles at intermediate 
radii.  Finally, a constant offset was subtracted across the entire 
spectrum.  This sky continuum level was determined to be that offset 
which results in the best fit to a de Vaucouleurs profile extending to 
large radii.  In practice little difference was found between the best
fit offset in the two wavelength ranges used to produce radial intensity
profiles. 

Extraction and dispersion correction of the spectra used standard IRAF
packages.  The input spectra to these tasks incorporate the relative and
absolute sky continuum adjustments described above, and have been median
combined with cosmic ray rejection as above, but have not been
rectified.  Instead, the extraction procedure traces the object spectrum
across the CCD. The galaxy spectra were extracted in apertures extending to 
$r_e/8$ from the galaxy center, as per G93.  Effective radii were taken 
from \citet{ch1faber89}.  
Galaxies were dispersion corrected using lamp spectra taken immediately
before or after the galaxy observations, at the same telescope position.
These lamp spectra were extracted using the same trace as the galaxy
spectra.  A high--order polynomial (typically fifth order) was fit to
the $\sim30$ lamp lines, and this dispersion correction was then applied to
the galaxy spectrum.  Typical wavelength calibration uncertainties are
$\sim 0.10\mbox{\AA}$.

\subsection{Lick Indices}

Line widths of the Lick indices defined in \citet{ch1trager98} and 
\citet{ch1wo97} were calculated using a version of the {\tt bwid} program,
provided by R.~M.~Rich \citep{ch1bwid}.  As {\tt bwid}
is not an IRAF procedure, the data were read out into ascii files for
input.  Test spectra from Guy Worthey's web page\footnote{ 
http://astro.wsu.edu/worthey/html/system.html}
were used to confirm the accuracy of {\tt bwid}.  Two sets of Lick/IDS indices 
were measured.  The primary set of indices consists of H$\beta$, Mg{\it b}, 
Fe5270, and Fe5335, while H$\gamma_{\rm F}$, Ca4227, G4300, Fe4383,
Ca4455, Fe4531, ${\rm C}_{2}4668$, Fe5015, and Fe5406 form the secondary set
of indices (Table~\ref{index1}).  The two sets are distinguished by both 
scientific and practical considerations.  The primary set comprises the 
indices in common use in 
previous observational studies \citep[e.~g.][]{ch1tfwg1}.  Furthermore,
the spectral range of the models of \citet{ch1schiavon} allow for sophisticated
error simulations of the primary indices, while simpler error estimates
must be used for many of the secondary indices; see \S~2.5 for details.

Several corrections must be applied before equivalent widths can be measured.
First, the galaxy sample must be adjusted to a common radial velocity, $v = 0$.
The extensive model library of \citet{ch1schiavon} was used to provide an
empirical Doppler correction.  The galaxy spectra were cross correlated with 
an old, metal rich model spectrum, and the resulting velocity was used in the
Doppler correction.

Next, the galaxy sample was corrected to zero velocity dispersion.
The following procedure was used.  The model spectra were rebinned to match the 
Doppler-corrected galaxy spectrum.  An analytic continuum function was fit to 
the galaxy and each model spectrum.  The model spectra were divided by their 
respective continuum functions, then multiplied by the galaxy continuum 
function to match the shape of the observed galaxy spectrum.  A copy of each 
model spectrum was then smoothed to the velocity dispersion of the galaxy
using the $\sigma$ measurements from \citet{ch1faber89}.  A least-squares fit 
was performed to select the appropriate model for each galaxy.  The fit
was performed within a spectral region surrounding the Mg~$b$, Fe5270, and
Fe5335 lines: $5140$--$5370\mbox{\AA}$.  
To empirically refine the galaxy's velocity dispersion, the $\sigma = 0$
version of the best fitting model spectrum was used as a template for the 
Pixfit code from \citet{ch1pixfit}.  The 
resulting $\sigma$ was then corrected for the different instrumental
resolution of the galaxy and model spectra.  The best fit model was then
rebinned to a logarithmic wavelength scale and
gaussian smoothed to the measured velocity dispersion of the galaxy, again
taking the intrinsic resolution of the model spectrum into account.
After transforming back to a linear wavelength scale,
line widths were measured in the $\sigma = 0$ model spectrum smoothed to
Lick/IDS resolution and the model spectrum smoothed to the 
galaxy's velocity dispersion and then to the Lick/IDS resolution.  The ratio 
of the two measurements is the correction factor applied to the corresponding 
equivalent width measured from the galaxy spectrum.  For a few indices
(Fe4383, Ca4455, Fe4531, and ${\rm C}_{2}4668$) no model spectra were 
available covering the appropriate wavelengths; for these indices 
$\sigma$-corrections were calculated from \citet{ch1trager98}.

As indicated above, it is necessary to degrade spectra to the resolution
of the Lick Image Dissector Scanner instrument which was used to make the
stellar observations upon which the Lick/IDS system is based.  The 
resolution function of this instrument is taken from \citet{ch1wo97}.  Since
the resolution varies strongly with wavelength at the blue end, smoothing was 
done piecewise in $100\mbox{\AA}$ increments in that part of the spectrum, 
while a constant value was appropriate for the red end.

An emission correction was required to accurately measure the H$\beta$, 
H$\gamma_{\rm F}$, and Fe5015 indices.  The
[OIII]$\lambda{5007}$ line was used to estimate the amount of H$\beta$ emission
(Trager et~al. 2000a).  Since [OIII]$\lambda{5007}$ is adjacent to the 
Fe$\lambda{5015}$ line, it is impossible to accurately determine the continuum
level.  Instead, model spectra were used.
The best fitting model was subtracted from the galaxy spectrum, and the
intensity of [OIII]$\lambda{5007}$ was measured on the difference spectrum.
This line was measured by hand using the IRAF task {\sc splot}.  
The result was then converted to equivalent width using an estimate of the
intensity level of the galaxy spectrum near [OIII]$\lambda{5007}$.  The
[OIII] emission was converted to an H$\beta$ correction using Equation~2 of
\citet{ch1tfwg1}.  H$\gamma_{\rm F}$ and Fe5015 were corrected as described
in \citet{ch1kuntschner}, with correction factors of 
$0.36\times{\rm [OIII]}\lambda{5007}$ and $0.61\times{\rm [OIII]}\lambda{5007}$ 
respectively.  In NGC~1052 H$\beta$ appears only
in emission.  NGC~3226, with the next strongest [OIII]$\lambda{5007}$ line,
also shows H$\beta$ emission within the aborption line, while all other
galaxies show at most a slight distortion in the shape of the H$\beta$
absorption.

Lick index standard stars were used to calibrate the output line widths.  
The reduction procedure for standard star observations is a simplified
version of that described above for the galaxy observations.  A typical
standard star is observed in five consecutive exposures of a few seconds
integration time.  After flat fielding and dark subtraction, the individual
frames are median combined and then extracted.  Sky subtraction is
performed via a linear fit between two sky regions near the star on the slit.
The stellar spectrum is extracted in a user-defined aperture extending
to the points at which the PSF disappears into the sky background.  This
aperture is traced across the CCD as above.  As
with the galaxy spectra, the accompanying calibration lamp spectra are 
extracted in the exact same aperture as the stellar spectrum.  After
dispersion calibration, the smoothing described above is applied. 
Radial velocities are determined empirically by cross correlation, in 
this case using the Jones spectra of the same star as templates.
The standard star spectrum is read out into an
ascii file, and line widths are measured by {\tt bwid}.
Corrections for each of the Lick indices are derived by comparing the 
observed stellar line widths for each observing run to the standard values.  
These calibration offsets are not constant from run to run, particularly
for the Fe5335 index.

The data from this project are not flux-calibrated, while the Lick/IDS
system was calibrated to a quartz lamp response curve.
This mismatch is absorbed into the calibration corrections.
The standard stars and galaxy spectra were observed with the same CCD, grating
and grating tilt, and are thus on a common system.
Errors in the mean for the IDS calibration to the primary indices
are $0.04$--$0.06\mbox{\AA}$, 
though run to run variations are often much larger.  Fully corrected 
measurements for each index, including estimates of the total error (\S~2.7) in
each index, are presented in Table~\ref{index1}.

Figure~\ref{index} shows the measurements for each index plotted against
the galaxy's velocity dispersion.  The Mg-$\sigma$ relation \citep{mgsig}
is clearly evident, with Ca4227 and ${\rm C}_{2}4668$ following the same 
trend.  The iron indices show similar behavior but with greater scatter, 
forming a mass-metallicity relation.  The two Balmer indices both follow the 
well-known trend that larger galaxies tend to have smaller values than smaller 
galaxies.  
The five galaxies with large ($>0.5\mbox{\AA}$) [OIII]$\lambda{5007}$ emission 
are omitted from subsequent analysis due to their large uncertainties.

\subsection{Error Simulations}

Empirical error simulations were performed as follows using the model spectra
described previously.  The model spectrum used in the velocity dispersion
correction forms the basis for the error simulations for the corresponding
galaxy.

Several potential sources of random error were simulated.  Poisson
noise was modeled by adding the quadrature sum of the galaxy poisson
noise and sky spectrum poisson noise to the model spectrum.  The poisson errors
were derived from the raw spectra (flat fielded only, no dark or sky
subtraction), in units of electrons rather than ADU.

Error in the subtraction of the night sky lines was estimated as follows.
The pixel value of the sky lines was multiplied by a small factor $f$ and
added in quadrature to the above poisson error.  The value of $f$ was
chosen such that the simulated spectrum best matched the true galaxy
spectrum.  Sky line subtraction error proved to be unimportant: $f = 0$
provides a good match to the galaxy spectra.

To investigate the possibility of flat fielding error, a noise floor was
introduced into the error calculations.  This floor was simply added in
quadrature to the poisson error, and the value of the noise floor was chosen
so that the simulated spectrum best matched the galaxy spectrum.  As with
sky line subtraction noise, flat fielding noise proved to be unimportant,
and a noise floor of zero was used.

Two other sources of error are directly measured and included in the
simulations.  The uncertainty in the dispersion solution is calculated by
IRAF, and the error in sky continuum subtraction is
easily estimated.  These errors are included in the simulations, the
first as a random shift in the wavelength scale and the second as a 
random shift in the continuum level of the model spectrum.  Sky subtraction
error may be underestimated by this method, particularly at the blue end
of the spectrum, far from the wavelength range in which the sky subtraction
measurements were made.

Error simulations were performed in groups of 1000 to provide accurate
statistics.  Each simulation spectrum was measured by {\tt bwid} in exactly
the same way that actual galaxy spectra were measured.  Three galaxies were 
used as the basis for these simulations:
NGC~3115, the galaxy with the highest $S/N$ in the sample, NGC~1209, a
galaxy with $S/N$ typical for the bulk of the galaxy sample, and NGC~4168,
a galaxy with relatively poor $S/N$.  

\subsection{Calibration Uncertainty}

The uncertainty in the calibration to the Lick/IDS system deserves
particular attention because for many indices this is the dominant source
of error.  The observed standard star primary index measurements are compared 
to the Lick/IDS values in Figs.~\ref{stdpre} and \ref{stdtpre}.  The former
shows that a constant calibration offset can correct the observations onto
the standard system --- no first-order term in index strength is needed for
the calibration.  Fig.~\ref{stdtpre} highlights the run-to-run variations
in the necessary calibration offset.  These offsets are calculated 
separately for each run, except for 2003 January 2 (Run~4) which was 
incorporated in the 2002 November (Run~3) run calibration due to the 
limited data obtained that
night.  The calibrated standard star index measurements are compared to the
Lick/IDS values in Figs.~\ref{stdpost} and \ref{stdtpost}, plotting
the difference between calibrated and standard values against the standard
value and the run number, respectively.  The external error in calibrating
the data to the Lick/IDS system is taken to be the error in the mean 
calibration offset for each run.  These external errors are shown at the
right of Figs.~\ref{stdtpre} and \ref{stdtpost}; the error bars are omitted
in Figs.~~\ref{stdpre} and \ref{stdpost} for clarity.

The repeated observations of a subset of standard stars in several observing
runs allows the internal calibration error to be measured.  Fig.~\ref{stdint}
shows these repeat observations for each run.  Instead of plotting the
difference between the calibrated and standard index values for each star
as above, in this figure the difference between calibrated individual 
observations and the average of all observations of that star is shown.  

Though not often listed, it is straightforward to calculate the uncertainty
within the Lick/IDS system in any given observing run.  \citet{ch1worthey94}
lists the typical error for an individual IDS measurement of each Lick
index.  Since each standard star was given equal weight in the calibration 
to the Lick/IDS system, the IDS internal error was calculated as the average 
of the errors in the individual stars, weighted by the number of IDS 
observations.

The calibration to the Lick/IDS system is shown in Fig.~\ref{stdchk}.  The
calibrated standard star observations from this work are in good agreement
with both the Lick/IDS system and standard star observations from G93.

\subsection{Additional Errors}

Several sources of error not included in the simulations
can also affect the equivalent width measurements
to varying degrees.  These include incorrect radial velocities, velocity
dispersions, emission correction, and calibration error, as well as possible 
problems with the adopted data reduction procedures.

Preliminary data reduction demonstrated that published radial velocities
for the galaxies or standard stars do not necessarily agree with the
observed spectra.  The mismatches could be over $100$~km/s in magnitude,
with correspondingly large effects on the equivalent width measurements.
This source of systematic error is avoided through the use of our empirical
radial velocity measurements, as described above.  Though the preliminary
reductions did not show any similar problem with published velocity 
dispersions, empirical measurements were again chosen for the same reason.

Velocity dispersion corrections are known to be very index-dependent 
\citep{ch1trager98}.  The fitting functions from that work were used to 
translate a conservative error estimate ${\sigma}_{\sigma} \approx 10$~km/s 
into index errors.  These errors proved to be negligible for H$\beta$
and Mg$b$.  The actual errors in $\sigma$ are estimated to be $\sim3$--5~km/s
by the Pixfit code.  
The uncertainty in the measured velocity dispersion is the quadrature sum
of the internal error measured by Pixfit with an estimate of the additional
uncertainty due to template mismatch.  The latter quantity was estimated
to be $\sim 3$~km/s based on comparing the measured velocity dispersions
using several model spectra which fit the data nearly as well as the
best fit spectrum.  The measured velocity dispersions
can easily be compared with data in the literature, as shown in 
Fig.~\ref{vdcomp}.  The mean offset is 3~km/s with an RMS scatter of 11~km/s.

The emission correction for H$\beta$, H$\gamma$, and Fe5015
can drastically increase the error for those galaxies for which the
correction is large.  The error in [OIII]$\lambda{5007}$ emission was estimated
by measuring the emission line in the difference spectrum (see above) using
several different continuum levels.  The average difference between these
alternate continuum level equivalent width measurements and the zero
continuum equivalent width measurement (which assumes a perfect match was
achieved between the continuum levels of the model and observed spectra)
was taken to be the uncertainty in the [OIII]$\lambda{5007}$ measurement.
For many galaxies with little or no [OIII]$\lambda{5007}$ emission, this
uncertainty is negligible, while for the galaxies with the largest 
[OIII]$\lambda{5007}$ emission lines the uncertainty is so large as to 
render the emission corrections extremely suspect.  The five galaxies 
with [OIII]$\lambda{5007}$ emission greater than $0.5\mbox{\AA}$ equivalent 
width have been omitted from subsequent analysis of the volume-limited
sample due to the large uncertainties.

An important potential source of systematic error is
the method of background subtraction.  Preliminary reductions tested two
different techniques.  The method using sky exposures is described above.
A simpler technique of simply subtracting a linear interpolation
of the background between two user-defined regions near the edge of the slit
was also tested.  The latter technique is the most common in the literature
\citep[G93 long slit observations,][]{beuing02}, though the possibility 
of age and abundance gradients makes it dubious for galaxies of large angular 
size.  For most galaxies in this study, the slit edges provide adequate
sky subtraction.  Reducing the same galaxy using both sky subtraction
methods results in final index measurements which differ by much less
than the total uncertainties.  This check also indicates that the 
technique described above using the sky exposures does not introduce 
significant wavelength-dependent biases into the spectrum at the blue end.

In addition to sky continuum subtraction, the bright Hg4358 night sky
line contaminates the G4300 and H$\gamma_{\rm F}$ indices.  The uncertainties
of those indices are undoubtedly larger than estimated, and the uncertainties
are likely to vary with galaxy redshift depending on whether the sky line
falls in the index bandpass or a sideband.

If the input images have non-negligible subpixel offsets from one another
in the dispersion direction, it is possible that the effective instrumental 
resolution could be artificially degraded when the images are combined into 
the final two dimensional galaxy spectrum.  To test for this,
one galaxy (NGC~1209) was re-reduced in a different manner following sky
subtraction.  Each input frame was reduced separately.  Wavelength calibration 
was performed on the two dimensional images using standard techniques
({\sc identify, reidentify, fitcoords}, and {\sc transform} in IRAF's 
{\sc longslit} package).
The $r_e/8$ aperture on each was traced and extracted, then doppler corrected
and smoothed to Lick/IDS resolution.  The average index values showed no
significant difference from those derived using the previous data reduction
technique.  Pixfit measured velocity dispersions less than $5$~km/s smaller
than those measured on the combined extraction previously.  We conclude that
there is little loss of effective instrumental resolution from combining input
images.

Since NGC~1209 was also used in the error simulations, the alternate
reduction procedure above also provides an independent test of the estimated
errors for each index.  The alternate technique should measure only the
random poisson errors between the spectra.  Thus {\it a priori} one expects
smaller error bars than those estimated from the simulations which
include several possible sources of systematic error as well.  The error
comparison confirmed this expectation, except for Mg$b$ where the two error
estimates were equal.

The final error bars adopted for the primary indices for each galaxy are a 
quadrature sum of the
errors from the error simulation appropriate for that galaxy's $S/N$, the errors
in $\sigma$-correction and emission correction, and the internal and
external calibration
errors.  Table~\ref{error} lists typical values for each of these quantities, as
well as the uncertainty within the IDS system for a typical observing run.
Errors for the secondary indices are taken from the scatter in
the measurements of each index in the NGC~1209 individual frames, added
in quadrature as above with the $\sigma$-correction, emission correction, and
IDS calibration errors.  Since the calibration error is a
large contribution to many of these index errors, the results should be
reasonably accurate even for galaxies of significantly lower $S/N$ than
NGC~1209.

As a final consistency check, index measurements from this study are
compared with measurements of the same galaxies from G93 and \citet{ch1denicolo}
in Fig.~\ref{comp}.  An equivalent comparison between G93 and Denicolo~et~al.
is shown in Fig.~\ref{gdcomp}.  The measurements from this study are found
to be in good agreement with G93.  The Denicolo~et~al. sample is found to
have larger error bars and systematic offsets in each measured quantity
compared to both G93 and this study.  Although crucial as an intermediate
step in establishing the consistency between the latter two data sets, the
Denicolo~et~al. sample will not be included in the discussion in \S~3.

\section{Analysis}

\subsection{Stellar Population Parameters}

The Balmer indices (H$\beta$, H$\gamma_{\rm F}$, and H$\delta_{\rm F}$)
are the most commonly used age indicators in stellar population analysis.
The data in this study do not include the H$\delta_{\rm F}$ index, but
both H$\beta$ and H$\gamma_{\rm F}$ have been measured.  \citet{ch1kuntschner}
presents a detailed discussion of the relative merits of each index
(see also \citet{crc03} for a discussion of the H$\beta$ index compared 
to higher order Balmer line ratios).
Briefly, H$\beta$ suffers from nearly twice the contamination of nebular
emission lines (if present) filling in the stellar absorption.  
H$\gamma_{\rm F}$ also offers increased dynamic range between young and
old stellar population models.  Offsetting this, H$\beta$ measurements
normally achieve higher $S/N$ due to the combination of the red colors
of early-type galaxies and the typical sensitivity curves of CCDs 
and spectrograph optics diminishing at bluer wavelengths.  Both indices 
can suffer from metal
line contamination in the continuum measurement.  \citet{ch1strader04} 
showed that H$\beta$ suffers from additional variance caused by weak Fe~I
lines in the wavelength range of the index.  As a result, ages based on
H$\beta$ can be less reliable than those based on H$\gamma$ or H$\delta$,
as seen in the extragalactic globular cluster study of \citet{ch1puzia}
(see also \citet{3610}).  However, H$\gamma_{\rm F}$ can suffer from CH
contamination and as noted in \S~2.7 is often affected by a bright night
sky line, neither of which are issues for H$\beta$.  Pragmatically, 
H$\beta$ is the age indicator of choice for this study, as the G93 data 
set does not include measurements of H$\gamma$ or H$\delta$.  A comparison
of age measurements using H$\beta$ and H$\gamma_{\rm F}$ for the galaxies
from this study will be presented below (\S~3.2).

The fully calibrated and corrected equivalent widths for the galaxy sample
are presented in Fig.~\ref{grid}, plotted on the H$\beta$--[MgFe]$^\prime$ 
plane.  The age--metallicity grid of the TMB models 
is also plotted.  The composite index [MgFe]$^\prime$ was defined in
TMB as an overall (i.e. [Z/H]) metallicity index that should be independent of
$[\alpha/{\rm Fe}]$ based on their calculations using the response
functions of \citet{ch1tb95}.  Members of the volume-limited sample observed
by G93 are also plotted.  The SSP age and metallicity for each galaxy was
measured by linear interpolation among the TMB models in each quantity.  

The $\alpha$-enhancement for each galaxy was measured using the TMB
models.  Since age does not strongly affect the location of lines of constant
$[\alpha/{\rm Fe}]$ in the Mg$b$--$\langle{\rm Fe}\rangle$ plane, each galaxy need not
be compared individually to the model of precisely matching age as measured
from Fig.~\ref{grid}.  The 3~Gyr model was used to measure $[\alpha/{\rm Fe}]$
for each galaxy, and an age correction was then applied.
This age correction was calculated by linearly 
interpolating the offsets perpendicular to isoenhancement lines between 
models with ages of 3, 8, and 15~Gyr.
The galaxy sample and the models are plotted in Fig.~\ref{alpha}.

Errors in age and metallicity are derived from the index errors in
H$\beta$ and [MgFe]$^\prime$.  The extrema of the error ellipse defined by 
$\sigma_{{\rm H}\beta}$ and $\sigma_{{\rm [MgFe]}^{\prime}}$
in the directions perpendicular to isochrone and isometallicity lines
were used to define $\sigma_t$ and $\sigma_{\rm Z}$ respectively.  These
stellar population uncertainties are highly correlated.  Errors 
introduced by the interpolation of the TMB model are estimated to 
be $0.05$--$0.1$~Gyr (depending on age) and $\le 0.01$~dex in [Z/H], in 
each case negligible compared to the errors derived from the index 
uncertainties.

The error in $\alpha$-enhancement is derived in a similar fashion, with the
added complication of including $\sigma_t$.  The extrema of the
$\sigma_{{\rm Mg}b}$--$\sigma_{\langle{\rm Fe}\rangle}$ error ellipse in the 
direction perpendicular to the isoenhancement lines are combined
with ages of $t \pm \sigma_t$ in the sense that maximizes the resulting
spread in measured $[\alpha/{\rm Fe}]$.

Parameter values and associated errors are presented in Table~\ref{ssp}.
Galaxies from G93 are
listed in the bottom portion of the table.  For consistency, the index
measurements from G93 were used to determine stellar population parameters
from TMB rather than using the parameters from \citet{ch1tfwg1}.

\subsection{Model Uncertainties}

The choice of model can strongly affect the values of the
physical quantities one derives from spectral line index measurements.
The importance of measuring the G93 galaxies with respect to the TMB models
is clearly shown in Figure~\ref{diff}.  The TMB models result in higher 
$[\alpha/{\rm Fe}]$ (by an average of 0.06~dex) and [Z/H] (by an average 
of 0.1~dex) than the Worthey models used by TFWG.  This suggests that the 
earlier models yield the correct Fe abundance, but underestimate the 
$\alpha$-elements and thus also [Z/H].

The H$\gamma$ index is measured in the data from this work,
and a comparison between the age estimates using H$\beta$ and H$\gamma$
can be made as a check on the extent to which the choice of Balmer index
affects the resulting age estimates.  Within this
Balmer line comparison, it is also instructive to compare the effect of 
using different SSP models.  The models of \citet{ch1schiavon} are used
for this purpose.  First, the Balmer index measurements are compared with
one another in Fig.~\ref{balmercomp}.  With the exception of four outliers,
the two indices are tightly correlated as expected.
Figures~\ref{hbfe}~\&~\ref{hgfe} show the galaxies
presented in this study plotted on the H$\beta$ vs. $\langle{\rm Fe}\rangle$ and
H$\gamma_{\rm F}$ vs. $\langle{\rm Fe}\rangle$ planes respectively.  The TMB and
Schiavon model grids are plotted on the former, and the \citet*{ch1tmk04}
and Schiavon model grids are plotted on the latter.  All models have 
$[\alpha/{\rm Fe}]=0$.  Ages derived using each set of models are
plotted against each other in Fig.~\ref{betagamma}.  As is apparent from
a visual inspection of the grids, an age offset of several Gyr exists
between the Thomas~et~al. and Schiavon model grids, in the sense of the
Thomas~et~al. models yielding (on average) younger ages on H$\beta$ grids and
older ages on H$\gamma_{\rm F}$ grids compared to the Schiavon models.
The outliers in Fig.~\ref{balmercomp} remain outliers in Fig.~\ref{betagamma}
using either set of models.  The quantitative difference between age
estimates from the two sets of models suggests that the choice of SSP 
model is more important than the choice of Balmer index when attempting
to measure ages by this method.  
Since the two sets of models each produce consistent age measurements for 
some range of ages (Thomas~et~al. models 
are consistent for old galaxies; Schiavon models are consistent for young 
galaxies), for the purposes of this study there is no clear reason to prefer 
one to the other.  Therefore the choice to use the TMB models and measure SSP
parameters on model grids with H$\beta$ as the age-sensitive index does
not introduce any additional systematic error due to choice of models or
Balmer index.  Note that the above comparisons do indicate the
presence of systematic errors arising from both sources, however the same
would be true regardless of the models or Balmer index selected.
Also, an inconsistency between ages derived using H$\beta$ and H$\gamma$ does
not necessarily imply that the SSP model being used is incorrect in some
way.  As noted in \S~1, single stellar populations are a necessary simplifying
assumption, but galaxies in general will have stellar populations of
varying ages and metallicities.  Such a composite stellar population can
easily yield different ages derived using H$\gamma$ as the age sensitive
index compared to using H$\beta$ as the age sensitive index.

\subsection{Multiple Stellar Population Analysis}

Current stellar population analysis techniques rely on the single stellar
population (SSP) assumption, that a single age, metallicity, and
$\alpha$-element abundance ratio can characterize the stellar population
of a galaxy.  More realistic stellar population
models can be constructed, however there is no widely applicable method
to fully constrain the problem --- even in the simplest case of assuming
two starbursts instead of one, the number of free parameters
is more than double that of the SSP case (two ages, two [Z/H], two
[$\alpha$/Fe], and the mass ratio between the two bursts).  Allowing
for a larger number of bursts or continuous star formation further
underconstrains the problem.  In a few galaxies, SSP observations of their
globular cluster (GC) populations (for which the SSP assumption is correct
rather than a means to obtain a luminosity-weighted average) may allow
two-burst multiple stellar population models.  Almost all elliptical
galaxies contain at least two populations of GCs, as seen in bimodal
color distributions \citep[e.~g.][]{zepf93}.  Measurements of these two
GC populations allow the two sets of stellar population parameters for a
two-burst galaxy model to be set in a consistent way.  \citet{3610} showed 
that this does not always work; if the youngest GC population is
coeval with the SSP age of the galaxy the resulting mass ratio between the
two starbursts will necessarily be strongly weighted towards the younger
burst.  As shown in \citet{ch1tfwg2}, for example, although a small mass
fraction of young stars is sufficient to change the measured SSP age by
several Gyr, mass fractions of 50\% or more are required to produce SSP
ages consistent with that of the young population itself.  Such large
burst fractions are contrary to most expectations for galaxy formation.

The GC populations of a few galaxies in the volume-limited sample have 
been studied spectroscopically.  With stellar population information for
two GC populations, the age, [Z/H], and [$\alpha$/Fe] for each burst can be 
estimated rather than being free parameters.  In practice, [$\alpha$/Fe] 
can be ignored as the best available models for multiple stellar population 
analysis remain those of \citet{w94model} via Worthey's Dial-a-Galaxy web 
page\footnote{ http://astro.wsu.edu/worthey/dial/dial\_a\_model.html}, which 
do not account for non-solar abundance ratios.

The requisite data exist for four galaxies: NGC~3115, NGC~3610, 
NGC~4365, and NGC~5846.  NGC~3610 is discussed in detail in \citet{3610}.  
Using IR photometry, \citet{4365ir} found candidate intermediate-age GCs 
associated with NGC~4365.  These candidates were seemingly confirmed by 
\citet{4365ref} finding a young and metal rich GC population with an age 
between 2--5~Gyr and approximately solar metallicity.  However,
\citet{brodie05} found that based on all Balmer lines the GC population
was coeval within the errors, with an age $>10$~Gyr.  The data from 
\citet{4365ref} were biased by low $S/N$ in the sense that some but not all 
of the intermediate age candidates from \citet{4365ir} appeared young using 
only the H$\beta$ index.  Since the GC population is older than the SSP age 
of the galaxy, no viable two burst model can be constructed using the GC 
populations to constrain the bursts.  NGC~3115 appears to be a similar case, 
with old GCs \citep{3115ref} around an intermediate age galaxy.  
\citet{ch1trager98} suggest that the galaxy itself is also old, however 
the NGC~3115 data presented here are of exemplary $S/N$ with negligible 
emission correction, and conclusively indicate an age between 4--7~Gyr 
depending on whether H$\beta$ or H$\gamma$ is used as the age-sensitive index.
The aperture used in this work is larger than that of the data analyzed 
in \citet{ch1trager98}, and \citet{bassin} showed that the contribution of 
a young stellar population increases with radius on the scale covered by 
aperture used here.  The discrepancy between the measurement presented here
and that of \citet{ch1trager98} is therefore explained by differing stellar 
populations in the apertures used in each study.
NGC~5846 \citep{puzia} has uniformly old GC 
populations at a range of metallicities.  Since the galaxy has an old SSP 
age (Table~\ref{ssp}), the SSP assumption appears to be a reasonable 
approximation to the star formation histories of this galaxy.

\subsection{Correlations Between SSP Parameters}

In Fig.~\ref{pc} all galaxies in the volume-limited sample observed thus 
far are plotted on the metallicity ``hyper-plane" of TFWG.  
Error ellipses in the hyper-plane projections were calculated from the
errors in the individual indices.  Index errors were treated as independent
quantities, as sky subtraction error is negligible at the centers of 
galaxies.  Monte Carlo simulations were run propagating index error
realizations for a particular galaxy into physical parameters and thence to 
the principal component axes of the hyper-plane.  The RMS scatter about
the major and minor axes of the Monte Carlo distribution in the hyper-plane
was used for the major and minor axis lengths of the $1 \sigma$ error
ellipses.  The actual error distribution is not quite 
symmetrical: as one expects from looking at the model grids, a fixed level 
of index scatter leads to unequal parameter scatter in the young and 
metal-rich directions compared to the old and metal-poor directions.
The hyper-plane is clearly due to physical variations between galaxies,
not measurement error or bias.  The volume-limited and G93 samples have
similar distributions within the hyper-plane.

The $Z$-plane from TFWG (Fig.~\ref{zplane}) provides an intriguing challenge
to theories of galaxy formation.  This projection shows that a linear 
relationship exists between [Z/H], log~$\sigma$, and log~$t$.  TFWG showed that 
no other galaxy parameters significantly affected this fit.  The fit from
TFWG is plotted; however a slight offset would better fit the data presented
here.  This is the result of using the TMB SSP models instead of the
Worthey models used by TFWG; recall that Fig.~\ref{diff} showed a constant
offset between the two sets of models in the [Z/H] measurements.  This
offset in [Z/H] requires a corresponding offset in the $Z$-plane fit.
As discussed in detail in TFWG, single burst models cannot plausibly produce 
a $Z$-plane 
relation --- a linear relation at the present epoch would be noticeably 
curved at past and future times, as old objects change less in log~$t$ than 
young objects.  TFWG showed that a two (or several) burst ``frosting" model 
can in principle maintain the $Z$-plane over time.  Such frosting populations 
most plausibly form from
gas within the galaxy and enriched by previous star-formation events; it is
unlikely that accreted gas or a merger event would have the appropriate 
metallicity to keep the galaxy on the $Z$-plane.  However, the young and 
metal-rich end of the $Z$-plane tends to be populated by likely merger 
remnants such as NGC~3610, the most extreme such galaxy in the 
volume-limited sample.  As shown in Fig.~\ref{zplane}, NGC~3610 lies 
directly on the $Z$-plane, as do other similar galaxies from the G93 sample
which are also believed to be merger remnants (TFWG).

The volume-limited sample is compared with the G93 sample in all four
principal component axes, as well as position along the edge-on projection
of the $Z$-plane.  K-S tests indicate that the volume-limited sample is
consistent with being drawn from the same distribution as the G93 sample
in all five dimensions.  This confirms that the G93 galaxies are 
representative of the early-type galaxy population in the local universe.

\subsection{Correlations with Other Parameters}

Having derived stellar population parameters for the galaxies in the
volume-limited sample, as well as rederiving the stellar population
parameters for the G93 sample using the same models, we now turn our
attention to other physical parameters of these galaxies.  Figure~\ref{asig}
shows the stellar population parameters age, [Z/H], and $[\alpha/{\rm Fe}]$
plotted against three structural parameters,
$\sigma$, $M_B$, and $r_e$, as well as the anisotropy parameter $(v/\sigma)_*$
\citep*{kappa, faber97}, the core profile slope $\gamma$ \citep{lauerprep}, 
and the $B-V$ color gradient \citep*{ch1idiart02}.
These quantities are listed in Table~\ref{supp}.
Note that the relatively small magnitude range covered by this sample
limits the ability to investigate correlations involving luminosity or
quantities closely tied to luminosity.
For each pair of parameters plotted against each other, the correlation
statistic was calculated, excluding those galaxies with large emission
corrections.  The Kendall's $\tau$ rank correlation test was 
used to identify correlations which were not significant.  Unfortunately
the Kendall's $\tau$ test can only rule out correlations; it cannot identify
which correlations are significant, as demonstrated in trials where it found
nominally significant correlations between NGC number and various physical
parameters.  The Spearman rank correlation test was found to have similar
behavior.  Thus a correlation is judged to be significant if it is not ruled
out by the Kendall's $\tau$ test and the absolute value of the correlation 
statistic is greater than a threshold value, arbitrarily chosen to be 0.3.
By these criteria, significant correlations are found between $\sigma$ and
both age and [$\alpha$/Fe]; and between log~$r_e$ and age.  The relation
between [$\alpha$/Fe] and $\sigma$ is a representation of the well-known 
Mg--$\sigma$ relation.  The other two relations indicate that the smallest 
galaxies, measured either by radius or mass, are younger than larger galaxies, 
as expected from the ``downsizing'' galaxy formation scheme of \citet{cowie}.
This result is in agreement with the more comprehensive study of low
velocity dispersion galaxies by \citet{crc03}.

Early-type galaxies lie on planes in two different parameter spaces,
the Fundamental Plane and the metallicity hyper-plane.  The existence of
a mapping from one plane to the other would imply a relation connecting
the size and brightness of a galaxy to its age and chemical composition.
To search for such a mapping, the $\kappa$-space parameterization of the
Fundamental Plane was used \citep{kappa}.  The two planes are shown
in Fig.~\ref{kappc}.  The IRAF task {\sc geomap}
found no combination of scale factors, rotations, and translations that
can map the $\kappa_1$--$\kappa_2$ face-on projection of the Fundamental
Plane into the PC1--PC2 face-on projection of the metallicity hyperplane.
The best fit resulted in average
offset distances of 0.15 in the PC1--PC2 plane (average offset of 0.07 in
the PC1 axis and 0.11 in the PC2 axis individually).  The largest offsets
occured for the most extreme galaxies in one or both axes, up to a
difference of 0.52 in PC2 for one galaxy.  As a result, the transformed
Fundamental Plane distribution covers a smaller portion of the PC1--PC2
plane than the metallicity hyper-plane.  The transformed Fundamental Plane
distribution is found to be consistent with the metallicity hyper-plane
by K-S tests, though the consistency is marginal (K-S probability of 8\%)
on the PC2 axis.  Also, as previously noted, the sample presented here
covers a relatively small range of luminosity and therefore may lack
leverage to reveal important details within these parameter spaces.

The stellar brightness profiles of 27 of the galaxies in both the
volume-limited sample and the G93 sample have been studied with HST
\citep{faber97, lauerprep}.  Figure~\ref{asig}
showed no significant correlations between the brightness profile slope 
$\gamma$ and any stellar population parameter, though this test was
restricted to the subset of the volume-limited sample for which HST
surface brightness profiles are available.  Combining all available galaxies,
including G93 galaxies not included in the volume-limited sample, the 
distribution of $\sigma$, age, [Z/H], and $[\alpha$/Fe] among both core 
and power-law galaxies was examined using K-S tests.  The two classes of 
galaxies are consistent with having the same distribution of age and 
metallicity (probabilities 0.087 and 0.82, respectively), while core 
galaxies have significantly higher velocity dispersions and $[\alpha$/Fe] 
values than power-law galaxies (probabilities of 0.019 and 0.036, 
respectively).  This is as expected given the result of
\citep{faber97} that the most luminous galaxies all have core profiles,
while all galaxies below a certain luminosity threshold ($M_V > -20.5$) 
have power-law profiles.  
To further explore the meaning of these relations, the distribution
of stellar population parameters between galaxies with boxy or disky
isophotes was examined.  Since \citet{faber97} found a strong correlation
between galaxies with disky isophotes and power law core profiles, the
expectation is that the disky galaxies will have lower $[\alpha$/Fe]
and $\sigma$ measurements.  Isophotal shape measurements for 15~galaxies
are listed in \citet{ch1idiart02}, and the $a_4/a$ parameter from
\citet{faber97} was used to determine the shapes of an additional 13
galaxies.  Boxy and disky galaxies are
consistent with being drawn from the same distribution in $\sigma$, age,
and [Z/H] (probabilities of 0.43, 0.23, and 0.75, respectively), while 
disky systems have significantly smaller $[\alpha$/Fe] than boxy systems 
as expected (probability of 0.048).

\citet{carollo97} and \citet{sc97} argue that 
more massive, spherical systems form stars earlier and
for a shorter duration than less massive, more flattened systems.  The
models of \citet{sc97} show that flattened systems lose metals more easily
due to supernova blow-out, which has the consequence of delaying metal
enrichment in the outer regions and thus delaying the peak of the star
formation rate.  This peak occurs when a metal-enriched multiphase ISM has
built up, at which time gas can cool (and thus form stars) much more 
efficiently.  By this
argument, flattened galaxies --- disky, power law profile, high
$(v/\sigma)_*$, low mass --- should have lower $[\alpha$/Fe] and younger ages than
boxy, core profile, low $(v/\sigma)_*$, high mass galaxies.  Many of these 
predictions are confirmed, as mentioned above: a significant correlation
is seen in the sense that smaller, less massive galaxies are younger than
larger, more massive galaxies (Fig.~\ref{asig});
galaxies with power-law profiles have significantly lower $[\alpha$/Fe]
than galaxies with core profiles; and galaxies with disky isophotes have
significantly lower $[\alpha$/Fe] than galaxies with boxy isophotes.  No
statistically significant difference in age distribution is seen between
core and power-law galaxies or boxy and disky galaxies, however, nor are
any correlations between stellar population parameters and $(v/\sigma)_*$ found
to be significant.

The models of \citet{sc97} treat galaxies as evolving and forming stars
in isolation.  As \citet{faber97} summarized, gas-rich merger events lead
naturally to disky galaxies with power-law profiles.  Simulations
\citep[e.~g.][]{bh96} show that gas can destroy box orbits and that the
gas (and thus star formation) in the remnant object is centrally concentrated.
However, \citet{faber97} also noted that the merger of the supermassive black
holes from the progenitor galaxies can produce core brightness profiles by
ejecting stars from the center of the merger remnant.
It is therefore unclear to what extent major mergers are expected to
yield correlations between structural and stellar population parameters.

To search for additional correlations, principal component analyses were
run on several combinations of parameters (Table~\ref{pca}).  The input
parameters were each reduced to zero mean and unit variance prior to the
analysis.  Combinations of parameters including luminosity ($M_V$), one 
structural variable ($\gamma$ or the break radius $r_b$), and one SSP 
variable were studied; in addition, all seven galaxy parameters were 
analyzed together.  Using the Kaiser criterion \citep{kaiser}, only those 
principal components which account for at least as much variance as a single 
normalized input variable are listed in Table~\ref{pca}.  No new planes
were discovered in which two orthogonal linear combinations of the input
variables could explain almost all of the variance in the three dimensional
input parameter space.  The distributions of galaxies in the parameter spaces
studied by this method are triaxial, in some cases close to prolate spheroids
and in other cases close to oblate spheroids.  

\section{Summary and Conclusions}

A new set of spectra of nearby elliptical galaxies has been obtained.  
Combining this new data set with a selected portion of the G93 sample, a
volume-limited sample of 45~galaxies has been defined, of which 35 have
been observed.  The ten missing galaxies have a distribution in luminosity
and velocity dispersion that is similar to the full set of 45 galaxies.

It is worth noting that there is no significant qualitative difference
between SSP models; for almost all galaxies the use of different models
produces only a systematic offset in stellar population parameters. 
Although different choices of models will result in different absolute 
SSP quantities, comparing one galaxy to another within the same set of 
models yields consistent results independent of the choice of models.  This
quantitative offset between different models is greatest when higher order 
Balmer lines such as H$\gamma$ are used as the age-sensitive index in 
age-metallicity grids.  The difference in ages derived using different 
sets of models is found to be greater than the difference in ages 
derived using different Balmer indices in model grids.  

A method to constrain the input parameters for a two-burst multiple stellar
population analysis has been proposed.  However, stellar population data are
available for the GC systems of only a few galaxies; four such galaxies are 
investigated here.  In none of these cases is a two-burst model based on GC 
age measurements viable.  Two galaxies are coeval with their GC populations, 
suggesting that SSP analysis is adequate for those galaxies.  The other two 
galaxies have SSP ages significantly younger than any of the associated GCs.

The existence of the metallicity hyper-plane and the $Z$-plane of TFWG is
confirmed by this new data set.  The distribution of galaxies within these
planes is very similar to that shown by TFWG, and this distribution is
shown to be due to physical variations well in excess of measurement
uncertainties.

The G93 galaxy sample analyzed by TFWG is representative of
the local early-type galaxy population.  This work is in good statistical
agreement with their results.

Comparisons between stellar population parameters and indicators of a galaxy's 
internal structure support the idea introduced by \citet{sc97} that size
and angular momentum (i.e. large and spherical compared to smaller and
flattened) are key distinctions between power law and
core profile galaxies. 
Disky, power-law profile galaxies are found to have significantly lower 
[$\alpha$/Fe] than boxy, core profile galaxies, though no significant trend 
in age is seen between core and power-law or boxy and disky galaxies.  
Age is found to decrease with size, measured either by radius or mass, as 
expected in downsizing models for galaxy formation.

\bigskip
\acknowledgments

This research has made use of the NASA/IPAC Extragalactic Database (NED) 
which is operated by the Jet Propulsion Laboratory, California Institute of 
Technology, under contract with the National Aeronautics and Space 
Administration.  This work also made use of the Gauss-Hermite Pixel
Fitting Software developed by R.P. van der Marel.  J.~H.~H. was supported 
in part by an ARCS Fellowship.  We would like to thank R.~Schiavon, S.~Faber,
P.~Guhathakurta, J.~Brodie, R.~Peterson, M.~Bolte, J.~Primack, M.~Geha, 
J.~Gonzalez, S.~Trager, and S. Thorsett for helpful discussions, and 
S.~Faber and T.~Lauer for allowing the use of their results in advance 
of publication.  We also thank the referee, Jim Rose, for his detailed
comments which greatly improved the final paper.

\vfill\eject

\clearpage
\begin{deluxetable}{lllll}
\tablecolumns{5}
\tablewidth{0pt}
\tablecaption{Volume Limited Galaxy Sample} 
\tablehead{
\colhead{Name} & \colhead{$\alpha$(J2000)} & \colhead{$\delta$(J2000)} & \colhead{$S/N$} & \colhead{Observed} \\}
\startdata
NGC 584 &   $01^{\rm h}31^{\rm m}20.7^{\rm s}$ &  $-06^{\circ}52^{\prime}06^{\prime\prime}$   & & G93  \\
NGC 596 &   $01^{\rm h}32^{\rm m}52.08^{\rm s}$ &  $-07^{\circ}01^{\prime}54.6^{\prime\prime}$   & 147.0 & 2002 November 3     \\
NGC 720 &   $01^{\rm h}53^{\rm m}00.4^{\rm s}$ &  $-13^{\circ}44^{\prime}18^{\prime\prime}$   & & G93  \\
NGC 821 &   $02^{\rm h}08^{\rm m}21.0^{\rm s}$ &  $+10^{\circ}59^{\prime}44^{\prime\prime}$   & & G93  \\
NGC 1052 &  $02^{\rm h}41^{\rm m}04.80^{\rm s}$ &  $-08^{\circ}15^{\prime}20.8^{\prime\prime}$  & 205.8 & 2000 November 26 \\
NGC 1172 &  $03^{\rm h}01^{\rm m}36.0^{\rm s}$  &    $-14^{\circ}50^{\prime}11^{\prime\prime}$   & 97.0 & 2002 November 3     \\
NGC 1199 &  $03^{\rm h}03^{\rm m}38.6^{\rm s}$ &  $-15^{\circ}36^{\prime}51^{\prime\prime}$      & & Not Observed        \\
NGC 1209 &  $03^{\rm h}06^{\rm m}03.0^{\rm s}$ &  $-15^{\circ}36^{\prime}40^{\prime\prime}$      & 149.2 & 2000 November 27     \\
NGC 1400 &  $03^{\rm h}39^{\rm m}31.0^{\rm s}$ &  $-18^{\circ}41^{\prime}22^{\prime\prime}$      & 121.0 & 2003 January 2    \\
NGC 1407 &  $03^{\rm h}40^{\rm m}11.8^{\rm s}$ &  $-18^{\circ}34^{\prime}48^{\prime\prime}$      & 164.3 & 2000 November 26     \\
NGC 2768 &   $09^{\rm h}11^{\rm m}37.50^{\rm s}$ &  $+60^{\circ}02^{\prime}15.0^{\prime\prime}$ & 89.7 & 2002 November 4       \\
NGC 2974 &   $09^{\rm h}42^{\rm m}32.96^{\rm s}$ &  $-03^{\circ}41^{\prime}55.2^{\prime\prime}$ & & Not Observed         \\
NGC 3115 & $10^{\rm h}05^{\rm m}13.42^{\rm s}$ &  $-07^{\circ}43^{\prime}06.5^{\prime\prime}$ & 305.0 & 2001 March 28 \\
NGC 3156 & $10^{\rm h}12^{\rm m}41.08^{\rm s}$ &  $+03^{\circ}07^{\prime}50.2^{\prime\prime}$  & & Not Observed   \\
NGC 3193 & $10^{\rm h}18^{\rm m}24.88^{\rm s}$ &  $+21^{\circ}53^{\prime}38.6^{\prime\prime}$     & 108.9 & 2004 March 27 \\
NGC 3226 & $10^{\rm h}23^{\rm m}27.00^{\rm s}$ &  $+19^{\circ}53^{\prime}54.4^{\prime\prime}$   & 101.9 & 2001 March 25 \\
NGC 3607 & $11^{\rm h}16^{\rm m}54.08^{\rm s}$ &  $+18^{\circ}03^{\prime}11.6^{\prime\prime}$  & 141.0 & 2003 January 2      \\
NGC 3608 & $11^{\rm h}16^{\rm m}58.7^{\rm s}$ &  $+18^{\circ}08^{\prime}57^{\prime\prime}$  & & G93    \\
NGC 3610 & $11^{\rm h}18^{\rm m}25.83^{\rm s}$ &  $+58^{\circ}47^{\prime}13.6^{\prime\prime}$  & 136.4 & 2001 March 27     \\
NGC 3613 & $11^{\rm h}18^{\rm m}36.12^{\rm s}$ &  $+58^{\circ}00^{\prime}04.5^{\prime\prime}$  & 107.4 & 2001 March 27       \\
NGC 3640 & $11^{\rm h}21^{\rm m}06.74^{\rm s}$ &  $+03^{\circ}14^{\prime}08.1^{\prime\prime}$  & 151.9 & 2001 March 28 \\
NGC 3962 & $11^{\rm h}54^{\rm m}40.0^{\rm s}$  &  $-13^{\circ}58^{\prime}30^{\prime\prime}$    & & Not Observed  \\
NGC 4125 & $12^{\rm h}08^{\rm m}05.71^{\rm s}$ &  $+65^{\circ}10^{\prime}24.5^{\prime\prime}$  & & Not Observed  \\
NGC 4168 & $12^{\rm h}12^{\rm m}16.9^{\rm s}$  &  $+13^{\circ}12^{\prime}20^{\prime\prime}$    & 103.7 & 2001 March 25 \\
NGC 4278 & $12^{\rm h}20^{\rm m}06.82^{\rm s}$ &  $+29^{\circ}16^{\prime}50.7^{\prime\prime}$  & & Not Observed  \\
NGC 4365 & $12^{\rm h}24^{\rm m}27.87^{\rm s}$ &  $+07^{\circ}19^{\prime}04.9^{\prime\prime}$  & 158.3 & 2001 March 26       \\
NGC 4374 & $12^{\rm h}25^{\rm m}03.7^{\rm s}$ &  $+12^{\circ}53^{\prime}14^{\prime\prime}$  & & G93    \\
NGC 4406 & $12^{\rm h}26^{\rm m}11.74^{\rm s}$ &  $+12^{\circ}56^{\prime}46.4^{\prime\prime}$  &  & Not Observed \\
NGC 4472 & $12^{\rm h}29^{\rm m}46.5^{\rm s}$ &  $+07^{\circ}59^{\prime}48^{\prime\prime}$  & & G93    \\
NGC 4473 & $12^{\rm h}29^{\rm m}48.87^{\rm s}$ &  $+13^{\circ}25^{\prime}45.7^{\prime\prime}$  & 186.7 & 2001 March 28 \\
NGC 4486 & $12^{\rm h}30^{\rm m}49.42^{\rm s}$ &  $+12^{\circ}23^{\prime}28.0^{\prime\prime}$  & 150.9 & 2004 March 27  \\
NGC 4552 & $12^{\rm h}35^{\rm m}39.9^{\rm s}$ &  $+12^{\circ}33^{\prime}55^{\prime\prime}$  & & G93    \\
NGC 4621 & $12^{\rm h}42^{\rm m}02.49^{\rm s}$ &  $+11^{\circ}38^{\prime}48.7^{\prime\prime}$  & 166.3 & 2004 March 27 \\
NGC 4636 & $12^{\rm h}42^{\rm m}49.7^{\rm s}$ &  $+02^{\circ}41^{\prime}18.4^{\prime\prime}$  & & Not Observed   \\
NGC 4649 & $12^{\rm h}43^{\rm m}39.7^{\rm s}$ &  $+11^{\circ}33^{\prime}09^{\prime\prime}$  & & G93    \\
NGC 4697 & $12^{\rm h}48^{\rm m}35.8^{\rm s}$ &  $-05^{\circ}48^{\prime}00^{\prime\prime}$  & & G93    \\
NGC 5322 & $13^{\rm h}49^{\rm m}15.19^{\rm s}$ &  $+60^{\circ}11^{\prime}26.2^{\prime\prime}$  & & Not Observed  \\
NGC 5485 & $14^{\rm h}07^{\rm m}11.5^{\rm s}$  &  $+55^{\circ}00^{\prime}07^{\prime\prime}$    & & Not Observed  \\
NGC 5576 & $14^{\rm h}21^{\rm m}04.11^{\rm s}$ &  $+03^{\circ}16^{\prime}13.5^{\prime\prime}$  & 144.0 & 2001 March 26  \\
NGC 5638 & $14^{\rm h}29^{\rm m}40.4^{\rm s}$ &  $+03^{\circ}14^{\prime}04^{\prime\prime}$  & & G93    \\
NGC 5812 & $15^{\rm h}00^{\rm m}57.0^{\rm s}$ &  $-07^{\circ}27^{\prime}19^{\prime\prime}$  & & G93    \\
NGC 5813 & $15^{\rm h}01^{\rm m}11.2^{\rm s}$ &  $+01^{\circ}42^{\prime}08^{\prime\prime}$  & & G93    \\
NGC 5831 & $15^{\rm h}04^{\rm m}07.2^{\rm s}$ &  $+01^{\circ}13^{\prime}15^{\prime\prime}$  & & G93    \\
NGC 5846 & $15^{\rm h}06^{\rm m}29.3^{\rm s}$ &  $+01^{\circ}36^{\prime}21^{\prime\prime}$  & 153.2 & G93, 2001 March 28         \\
NGC 6703 & $18^{\rm h}47^{\rm m}18.9^{\rm s}$ &  $+45^{\circ}33^{\prime}02^{\prime\prime}$  & & G93    \\
\enddata
\label{vlsample}
\end{deluxetable}

\begin{turnpage}
\begin{deluxetable}{lllrllllllllllllll}
\tabletypesize{\scriptsize}
\tablecolumns{18}
\tablewidth{0pt}
\tablecaption{Index Measurements: $r_e/8$ Aperture}
\tablehead{
\colhead{Galaxy} & \colhead{$\sigma$} & \colhead{Ca4227} & \colhead{G4300} & \colhead{H$\gamma_{\rm F}$} & \colhead{Fe4383} & \colhead{Ca4455} & \colhead{Fe4531} & \colhead{${\rm C}_24668$} & \colhead{{\bf H$\beta$}} & \colhead{[OIII]} & \colhead{Fe5015} & \colhead{Mg$_2$} & \colhead{{\bf Mg$b$}} &
\colhead{{\bf Fe5270}} & \colhead{{\bf Fe5335}} & \colhead{Fe5406} & \colhead{[MgFe]$^\prime$} \\
}
\startdata
NGC 596 & 151 & 1.19 & 5.33 & -1.12 & 4.54 & 2.06 & 3.62 & 6.60 & {\bf 1.94} & -0.06 & 5.43 & 0.260 & {\bf 3.89} & {\bf 2.91} & {\bf 2.46} & 1.72 & 3.29 \\
~~ & 4 & 0.10 & 0.14 & 0.13 & 0.21 & 0.14 & 0.09 & 0.20 & {\bf 0.06} & 0.03 & 0.28 & 0.003 & {\bf 0.05} & {\bf 0.06} & {\bf 0.09} & 0.09 & 0.05 \\
NGC 1052 & 215 & 1.20 & 5.72 & -2.17 & 6.50 & 1.43 & 3.78 & 8.24 & {\bf 1.21} &
-3.71 & 1.90 & 0.340 & {\bf 5.96} & {\bf 3.05} & {\bf 2.78} & 1.88 & 4.21 \\
~~ & 4 & 0.10 & 0.14 & 0.38 & 0.09 & 0.05 & 0.11 & 0.12 & {\bf 0.7} & 1.0 & 0.64
 & 0.003 & {\bf 0.05} & {\bf 0.06} & {\bf 0.09} & 0.09 & 0.05 \\
NGC 1172 & 113 & 1.21 & 5.29 & -1.16 & 4.38 & 1.35 & 3.36 & 4.97 & {\bf 1.94} &
-0.65 & 4.91 & 0.238 & {\bf 3.89} & {\bf 2.70} & {\bf 2.27} & 1.61 & 3.17 \\
~~ & 4 & 0.10 & 0.14 & 0.15 & 0.21 & 0.14 & 0.09 & 0.20 & {\bf 0.15} & 0.20 & 0.30 & 0.003 & {\bf 0.06} & {\bf 0.09} & {\bf 0.13} & 0.09 & 0.07 \\
NGC 1209 & 225 & 1.39 & 5.35 & -1.85 & 5.58 & 1.58 & 3.76 & 7.75 & {\bf 1.27} &
-0.23 & 5.91 & 0.326 & {\bf 4.99} & {\bf 3.32} & {\bf 2.78} & 1.72 & 3.98 \\
~~ & 4 & 0.10 & 0.14 & 0.12 & 0.09 & 0.05 & 0.11 & 0.12 & {\bf 0.12} & 0.15 & 0.22 & 0.003 & {\bf 0.05} & {\bf 0.06} & {\bf 0.09} & 0.09 & 0.05 \\
NGC 1400 & 285 & 1.39 & 5.08 & -1.80 & 4.72 & 1.47 & 3.90 & 7.47 & {\bf 1.33} &
-0.21 & 6.03 & 0.336 & {\bf 5.32} & {\bf 3.01} & {\bf 2.75} & 1.82 & 3.95 \\
~~ & 5 & 0.10 & 0.14 & 0.14 & 0.21 & 0.14 & 0.09 & 0.20 & {\bf 0.11} & 0.13 & 0.29 & 0.003 & {\bf 0.06} & {\bf 0.09} & {\bf 0.13} & 0.09 & 0.08 \\
NGC 1407 & 296 & 1.47 & 4.80 & -1.18 & 6.29 & 1.30 & 3.87 & 8.80 & {\bf 1.40} &
-0.01 & 5.90 & 0.350 & {\bf 5.41} & {\bf 3.50} & {\bf 2.83} & 1.86 & 4.23 \\
~~ & 4 & 0.10 & 0.14 & 0.11 & 0.09 & 0.05 & 0.11 & 0.12 & {\bf 0.06} & 0.02 & 0.20 & 0.003 & {\bf 0.05} & {\bf 0.06} & {\bf 0.09} & 0.09 & 0.05 \\
NGC 2768 & 211 & 1.46 & 5.36 & -1.45 & 5.36 & 1.50 & 3.80 & 5.23 & {\bf 1.62} &
-0.58 & 4.29 & 0.200 & {\bf 4.36} & {\bf 2.81} & {\bf 2.70} & 1.92 & 3.48 \\
~~ & 4 & 0.09 & 0.18 & 0.20 & 0.20 & 0.05 & 0.11 & 0.11 & {\bf 0.35} & 0.47 & 0.58 & 0.003 & {\bf 0.07} & {\bf 0.10} & {\bf 0.14} & 0.09 & 0.08 \\
NGC 3115 & 276 & 1.55 & 5.57 & -1.78 & 5.61 & 1.53 & 4.03 & 8.52 & {\bf 1.66} &
-0.03 & 6.23 & 0.325 & {\bf 5.04} & {\bf 3.33} & {\bf 3.21} & 2.07 & 4.08 \\
~~ & 4 & 0.10 & 0.14 & 0.11 & 0.09 & 0.05 & 0.11 & 0.12 & {\bf 0.04} & 0.01 & 0.20 & 0.003 & {\bf 0.05} & {\bf 0.05} & {\bf 0.07} & 0.09 & 0.04 \\
NGC 3193 & 228 & 1.27 & 5.17 & -1.64 & 5.10 & 1.59 & 3.94 & 7.52 & {\bf 1.50} &
-0.45 & 5.28 & 0.318 & {\bf 4.61} & {\bf 3.02} & {\bf 2.67} & 2.16 & 3.67 \\
~~ & 4 & 0.09 & 0.18 & 0.13 & 0.20 & 0.05 & 0.11 & 0.11 & {\bf 0.16} & 0.18 & 0.58 & 0.003 & {\bf 0.07} & {\bf 0.10} & {\bf 0.14} & 0.09 & 0.08 \\
NGC 3226 & 180 & 1.09 & 5.85 & -2.17 & 4.75 & 1.39 & 3.67 & 7.26 & {\bf 1.73} &
-1.48 & 4.14 & 0.294 & {\bf 4.73} & {\bf 2.78} & {\bf 2.77} & 1.78 & 3.62 \\
~~ & 5 & 0.10 & 0.16 & 0.12 & 0.18 & 0.07 & 0.07 & 0.19 & {\bf 0.17} & 0.23 & 0.18 & 0.003 & {\bf 0.06} & {\bf 0.09} & {\bf 0.13} & 0.09 & 0.08 \\
NGC 3607 & 231 & 1.14 & 5.27 & -1.53 & 4.86 & 1.47 & 3.78 & 8.22 & {\bf 1.53} &
-0.30 & 5.89 & 0.313 & {\bf 4.76} & {\bf 2.96} & {\bf 2.78} & 1.83 & 3.72 \\
~~ & 4 & 0.10 & 0.14 & 0.14 & 0.21 & 0.14 & 0.09 & 0.20 & {\bf 0.11} & 0.13 & 0.28 & 0.003 & {\bf 0.05} & {\bf 0.06} & {\bf 0.09} & 0.09 & 0.05 \\
NGC 3610 & 172 & 0.89 & 4.99 & -0.52 & 4.72 & 1.52 & 3.57 & 8.08 & {\bf 2.39} &
-0.03 & 5.51 & 0.255 & {\bf 4.04} & {\bf 2.77} & {\bf 3.10} & 1.81 & 3.40 \\
~~ & 4 & 0.10 & 0.16 & 0.09 & 0.18 & 0.07 & 0.07 & 0.19 & {\bf 0.08} & 0.08 & 0.12 & 0.003 & {\bf 0.05} & {\bf 0.06} & {\bf 0.09} & 0.09 & 0.05 \\
NGC 3613 & 221 & 1.29 & 5.16 & -1.46 & 5.00 & 1.77 & 3.35 & 7.43 & {\bf 1.85} &
0.00 & 5.92 & 0.289 & {\bf 4.47} & {\bf 3.21} & {\bf 3.15} & 2.00 & 3.78 \\
~~ & 4 & 0.10 & 0.16 & 0.08 & 0.18 & 0.07 & 0.07 & 0.19 & {\bf 0.06} & 0.01 & 0.11 & 0.003 & {\bf 0.06} & {\bf 0.09} & {\bf 0.13} & 0.09 & 0.07 \\
NGC 3640 & 178 & 1.20 & 5.22 & -1.29 & 4.88 & 1.51 & 3.61 & 6.83 & {\bf 1.87} &
-0.03 & 5.79 & 0.266 & {\bf 3.96} & {\bf 2.96} & {\bf 2.84} & 1.80 & 3.40 \\
~~ & 4 & 0.10 & 0.16 & 0.08 & 0.18 & 0.07 & 0.07 & 0.19 & {\bf 0.06} & 0.02 & 0.11 & 0.003 & {\bf 0.05} & {\bf 0.06} & {\bf 0.09} & 0.09 & 0.05 \\
NGC 4168 & 179 & 1.08 & 5.07 & -1.08 & 4.85 & 1.59 & 3.52 & 6.65 & {\bf 1.87} &
-0.26 & 5.06 & 0.258 & {\bf 3.98} & {\bf 2.92} & {\bf 2.69} & 1.61 & 3.37 \\
~~ & 5 & 0.10 & 0.16 & 0.08 & 0.18 & 0.07 & 0.07 & 0.19 & {\bf 0.07} & 0.05 & 0.11 & 0.003 & {\bf 0.06} & {\bf 0.09} & {\bf 0.13} & 0.09 & 0.07 \\
NGC 4365 & 270 & 1.58 & 5.41 & -1.48 & 5.24 & 1.95 & 4.04 & 8.78 & {\bf 1.55} &
-0.01 & 5.57 & 0.330 & {\bf 5.18} & {\bf 3.32} & {\bf 3.19} & 2.07 & 4.12 \\
~~ & 4 & 0.10 & 0.16 & 0.08 & 0.18 & 0.07 & 0.07 & 0.19 & {\bf 0.06} & 0.01 & 0.11 & 0.003 & {\bf 0.05} & {\bf 0.06} & {\bf 0.09} & 0.09 & 0.05 \\
NGC 4473 & 201 & 1.31 & 5.62 & -2.01 & 5.48 & 1.70 & 3.79 & 7.64 & {\bf 1.74} &
-0.01 & 5.73 & 0.315 & {\bf 4.73} & {\bf 3.22} & {\bf 3.02} & 1.81 & 3.87 \\
~~ & 4 & 0.10 & 0.16 & 0.08 & 0.18 & 0.07 & 0.07 & 0.19 & {\bf 0.06} & 0.01 & 0.11 & 0.003 & {\bf 0.05} & {\bf 0.06} & {\bf 0.09} & 0.09 & 0.05 \\
NGC 4486 & 371 & 1.71 & 5.47 & 1.37 & 5.71 & 1.30 & 4.24 & 8.71 & {\bf 1.07} & -.57 & 3.63 & 0.348 & {\bf 5.78} & {\bf 2.88} & {\bf 3.22} & 2.37 & 4.15 \\
~~ & 5 & 0.09 & 0.18 & 0.16 & 0.20 & 0.05 & 0.11 & 0.11 & {\bf 0.25} & 0.32 & 0.58 & 0.003 & {\bf 0.06} & {\bf 0.07} & {\bf 0.11} & 0.09 & 0.07 \\
NGC 4621 & 260 & 1.45 & 5.50 & -2.07 & 5.34 & 1.54 & 4.03 & 8.39 & {\bf 1.26} &
0.0 & 5.66 & 0.348 & {\bf 5.09} & {\bf 3.15} & {\bf 3.07} & 2.15 & 3.99 \\
~~ & 4 & 0.09 & 0.18 & 0.11 & 0.20 & 0.05 & 0.11 & 0.11 & {\bf 0.10} & 0.01 & 0.58 & 0.003 & {\bf 0.06} & {\bf 0.07} & {\bf 0.11} & 0.09 & 0.07 \\
NGC 5576 & 190 & 1.10 & 5.36 & -1.04 & 4.66 & 1.41 & 3.57 & 7.32 & {\bf 2.10} &
-0.01 & 5.52 & 0.271 & {\bf 4.13} & {\bf 3.03} & {\bf 2.83} & 1.79 & 3.50 \\
~~ & 4 & 0.10 & 0.16 & 0.08 & 0.18 & 0.07 & 0.07 & 0.19 & {\bf 0.06} & 0.02 & 0.11 & 0.003 & {\bf 0.05} & {\bf 0.06} & {\bf 0.09} & 0.09 & 0.05 \\
NGC 5846 & 243 & 1.40 & 5.09 & -1.16 & 5.39 & 1.51 & 3.70 & 8.12 & {\bf 1.44} &
-0.23 & 4.81 & 0.333 & {\bf 5.20} & {\bf 3.13} & {\bf 3.03} & 1.91 & 4.02 \\
~~ & 4 & 0.10 & 0.16 & 0.10 & 0.18 & 0.07 & 0.07 & 0.19 & {\bf 0.12} & 0.15 & 0.11 & 0.003 & {\bf 0.05} & {\bf 0.06} & {\bf 0.09} & 0.09 & 0.05 \\
\enddata
\label{index1}
\end{deluxetable}
\end{turnpage}

\begin{deluxetable}{lllll}
\tablecaption{Primary Index Error Table} 
\tablecolumns{5}
\tablewidth{0pt}
\tablehead{
\colhead{Error} & \colhead{H$\beta$} & \colhead{Mg$b$} & \colhead{Fe5270} & \colhead{Fe5335} \\
}
\startdata
Simulations & 0.05 & 0.03 & 0.04 & 0.06 \\
Velocity Dispersion & 0.00 & 0.00 & 0.01 & 0.02 \\
Calibration - External & 0.05 & 0.05 & 0.04 & 0.06 \\
Calibration - Internal & 0.02 & 0.02 & 0.02 & 0.04 \\
Lick/IDS System & 0.07 & 0.07 & 0.08 & 0.08 \\
\enddata
\label{error}
\end{deluxetable}

\begin{deluxetable}{lllrlllll}
\tablecolumns{9}
\tablewidth{0pt}
\tablecaption{Stellar Population Parameters} 
\tablehead{
\colhead{Galaxy} & \colhead{$\sigma$} & \colhead{$\sigma_{\sigma}$} & \colhead{Age} & \colhead{$\sigma_t$} & \colhead{[Z/H]} & \colhead{$\sigma_Z$} & \colhead{$[\alpha/{\rm Fe}]$} & \colhead{$\sigma_{\alpha}$} \\
	& (km/s) & (km/s) & (Gyr) & (Gyr) & (dex) & (dex) & (dex) & (dex) \\
}
\startdata
\multicolumn{9}{c}{This work, TMB models:}\\
NGC 0596 & 151 & 4 &  4.4 & 0.7 & 0.22 & 0.1  & 0.19 & 0.04 \\
NGC 1052 & 215 & 4 & 16 & 14 & 0.42 & 0.3 & 0.44 & 0.12 \\
NGC 1172 & 113 & 4 & 4.8  & 1.5 & 0.13 & 0.13 & 0.26 & 0.05 \\
NGC 1209 & 225 & 4 & 15.6 & 3.0 & 0.28 & 0.08 & 0.23 & 0.03 \\
NGC 1400 & 285 & 5 & 14.2 & 3.0 & 0.31 & 0.13 & 0.35 & 0.04 \\
NGC 1407 & 296 & 4 & 9.5  & 2.2 & 0.56 & 0.07 & 0.30 & 0.04 \\
NGC 2768 & 211 & 4 & 10 & 7.0 & 0.14 & 0.25 & 0.22 & 0.10 \\
NGC 3115 & 276 & 4 & 3.9  & 0.7 & 0.65 & 0.06 & 0.25 & 0.03 \\
NGC 3193 & 228 & 4 & 11.8 & 3.2 & 0.20 & 0.12 & 0.24 & 0.06 \\
NGC 3226 & 180 & 5 & 6.1  & 3.5 & 0.33 & 0.17 & 0.32 & 0.06 \\
NGC 3607 & 231 & 4 & 10.6 & 2.3 & 0.27 & 0.1  & 0.27 & 0.04 \\
NGC 3610 & 172 & 4 & 1.7  & 0.1 & 0.76 & 0.16 & 0.28 & 0.03 \\
NGC 3613 & 221 & 4 & 3.3  & 1.0 & 0.59 & 0.12 & 0.17 & 0.05 \\
NGC 3640 & 178 & 4 & 4.9  & 1.0 & 0.26 & 0.08 & 0.13 & 0.04 \\
NGC 4168 & 179 & 5 & 5.0  & 1.4 & 0.24 & 0.1  & 0.17 & 0.05 \\
NGC 4365 & 270 & 4 & 5.9  & 1.9 & 0.59 & 0.08 & 0.26 & 0.04 \\
NGC 4473 & 201 & 4 & 4.0  & 1.3 & 0.56 & 0.08 & 0.24 & 0.04 \\
NGC 4486 & 371 & 5 & 19.6 & 7.5 & 0.27 & 0.18 & 0.36 & 0.08 \\
NGC 4621 & 260 & 4 & 15.8 & 4.0 & 0.29 & 0.11 & 0.23 & 0.05 \\
NGC 5576 & 190 & 4 & 2.5  & 0.4 & 0.60 & 0.11 & 0.21 & 0.05 \\
NGC 5846 & 243 & 4 & 10.6 & 4.4 & 0.44 & 0.14 & 0.28 & 0.03 \\
\hline \\
\multicolumn{9}{c}{Galaxies from G93, volume-limited sample, TMB models:}\\
NGC 0584 & 236 & 3 & 2.4 & 0.3 & 0.61 & 0.04 & 0.24 & 0.03 \\
NGC 0720 & 239 & 5 & 3.7 & 1.8 & 0.55 & 0.09 & 0.40 & 0.08 \\
NGC 0821 & 189 & 3 & 7.3 & 1.5 & 0.33 & 0.02 & 0.22 & 0.03 \\
NGC 3608 & 178 & 3 & 6.1 & 1.5 & 0.38 & 0.04 & 0.24 & 0.04 \\
NGC 4374 & 282 & 3 & 11.1 & 1.5 & 0.24 & 0.02 & 0.28 & 0.02 \\
NGC 4472 & 279 & 4 & 7.8 & 1.5 & 0.36 & 0.04 & 0.29 & 0.03 \\
NGC 4552 & 252 & 3 & 10.5 & 1.5 & 0.36 & 0.03 & 0.31 & 0.02 \\
NGC 4649 & 310 & 3 & 11.9 & 1.5 & 0.37 & 0.03 & 0.32 & 0.02 \\
NGC 4697 & 162 & 4 & 7.1 & 1.8 & 0.19 & 0.04 & 0.18 & 0.03 \\
NGC 5638 & 154 & 3 & 7.8 & 1.5 & 0.32 & 0.03 & 0.26 & 0.03 \\
NGC 5812 & 200 & 3 & 5.0 & 1.1 & 0.47 & 0.03 & 0.26 & 0.03 \\
NGC 5813 & 205 & 3 & 14.9 & 2.3 & 0.07 & 0.04 & 0.29 & 0.03 \\
NGC 5831 & 161 & 3 & 2.7 & 0.2 & 0.61 & 0.04 & 0.20 & 0.03 \\
NGC 5846 & 224 & 4 & 12.2 & 2.4 & 0.25 & 0.04 & 0.29 & 0.03 \\
NGC 6703 & 183 & 3 & 3.9 & 1.0 & 0.39 & 0.05 & 0.21 & 0.03 \\
\hline \\
\multicolumn{9}{c}{Other galaxies from G93, TMB models:}\\
NGC 0221 & 72 & 3 & 2.8 & 0.7 & 0.10 & 0.05 & -0.07 & 0.01 \\
NGC 0315 & 321 & 4 & 5.0 & 1.5 & 0.44 & 0.06 & 0.32 & 0.02 \\
NGC 0507 & 262 & 6 & 6.9 & 2.8 & 0.29 & 0.07 & 0.27 & 0.03 \\
NGC 0547 & 236 & 4 & 8.3 & 2.4 & 0.34 & 0.05 & 0.33 & 0.02 \\
NGC 0636 & 160 & 3 & 3.8 & 0.7 & 0.44 & 0.07 & 0.18 & 0.02 \\
NGC 1453 & 286 & 4 & 7.1 & 1.9 & 0.42 & 0.06 & 0.29 & 0.02 \\
NGC 1600 & 315 & 4 & 7.6 & 2.2 & 0.47 & 0.06 & 0.30 & 0.02 \\
NGC 1700 & 227 & 3 & 2.3 & 0.7 & 0.63 & 0.10 & 0.17 & 0.03 \\
NGC 2300 & 252 & 3 & 5.5 & 1.5 & 0.48 & 0.05 & 0.32 & 0.02 \\
NGC 2778 & 154 & 3 & 5.0 & 1.8 & 0.40 & 0.09 & 0.30 & 0.03 \\
NGC 3377 & 108 & 3 & 3.5 & 0.8 & 0.30 & 0.06 & 0.27 & 0.02 \\
NGC 3379 & 203 & 3 & 8.0 & 1.4 & 0.32 & 0.03 & 0.28 & 0.01 \\
NGC 3818 & 173 & 4 & 5.2 & 1.8 & 0.47 & 0.08 & 0.30 & 0.03 \\
NGC 4261 & 288 & 3 & 14.5 & 3.3 & 0.29 & 0.04 & 0.27 & 0.01 \\
NGC 4478 & 128 & 2 & 4.3 & 2.3 & 0.40 & 0.10 & 0.22 & 0.03 \\
NGC 4489 & 47 & 4 & 2.3 & 0.4 & 0.24 & 0.06 & 0.04 & 0.02 \\
NGC 6127 & 239 & 4 & 10.8 & 2.2 & 0.28 & 0.04 & 0.30 & 0.02 \\
NGC 6702 & 174 & 3 & 1.4 & 0.1 & 0.80 & 0.07 & 0.16 & 0.03 \\
NGC 7052 & 274 & 4 & 11.7 & 3.1 & 0.27 & 0.05 & 0.31 & 0.02 \\
NGC 7454 & 106 & 3 & 4.7 & 1.0 & 0.04 & 0.04 & 0.13 & 0.02 \\
NGC 7562 & 248 & 3 & 7.1 & 1.6 & 0.31 & 0.04 & 0.24 & 0.01 \\
NGC 7619 & 300 & 3 & 13.5 & 2.2 & 0.31 & 0.03 & 0.25 & 0.01 \\
NGC 7626 & 253 & 3 & 12.0 & 2.4 & 0.27 & 0.03 & 0.32 & 0.01 \\
NGC 7785 & 240 & 3 & 7.9 & 2.3 & 0.31 & 0.05 & 0.23 & 0.02 \\
\enddata
\label{ssp}
\end{deluxetable}

\begin{deluxetable}{llclll}
\tablecolumns{6}
\tablewidth{0pt}
\tablecaption{Supplementary Parameters} 
\tablehead{
\colhead{Galaxy} & \colhead{$M_B$} & \colhead{log$r_e$ (pc)} & \colhead{$(v/\sigma)_*$} & \colhead{$\gamma$} & \colhead{$\Delta_{BV}$} \\
}
\startdata
\multicolumn{6}{c}{Galaxies from this work:}\\
NGC 0596 & -19.58 & 3.41 & 0.67 & 0.16 & --- \\
NGC 1052 & -20.11 & 3.53 & 1.00 & 0.18 & --- \\
NGC 1172 & -19.48 & 3.68 & --- & -0.01 & --- \\
NGC 1209 & -19.72 & 3.40 & --- & --- & --- \\
NGC 1400 & -20.16 & 3.62 & --- & -0.10 & --- \\
NGC 1407 & -21.21 & 3.90 & 0.84 & --- & --- \\
NGC 2768 & -20.26 & 3.65 & --- & 0.24 & -0.054 \\
NGC 3115 & -20.30 & 3.29 & 1.25 & 0.52 & -0.069 \\
NGC 3193 & -20.28 & 3.48 & 0.80 & 0.01 & -0.086\\
NGC 3226 & -20.16 & 3.83 & --- & 0.00 & --- \\
NGC 3607 & -21.24 & 3.86 & 0.92 & 0.26 & -0.067 \\
NGC 3610 & -20.28 & 3.19 & 1.10 & 0.76 & -0.075 \\
NGC 3613 & -19.99 & 3.44 & 0.84 & 0.04 & -0.074 \\
NGC 3640 & -19.99 & 3.63 & 1.48 & -0.10 & -0.043 \\
NGC 4168 & -20.18 & 3.77 & 0.22 & 0.17 & --- \\
NGC 4365 & -20.34 & 3.64 & 0.08 & 0.07 & -0.058 \\
NGC 4473 & -19.77 & 3.28 & 0.40 & -0.07 & -0.063 \\
NGC 4486 & -21.46 & 3.90 & 0.11 & 0.27 & -0.063 \\
NGC 4621 & -20.33 & 3.55 & 0.74 & 0.75 & -0.081 \\
NGC 5576 & -19.84 & 3.29 & 0.22 & 0.01 & -0.076 \\
NGC 5846 & -21.15 & 4.02 & 0.10 & --- & -0.051\\
\hline
\multicolumn{6}{c}{Galaxies from G93, volume-limited sample:}\\
NGC 0584 & -20.39 & 3.48 & 1.55 & -0.01 & --- \\
NGC 0720 & -21.16 & 3.75 & 0.32 & 0.06 & --- \\
NGC 0821 & -20.27 & 3.64 & 0.70 & 0.10 & --- \\
NGC 3608 & -20.19 & 3.61 & 0.27 & 0.09 & -0.064 \\
NGC 4374 & -21.39 & 3.68 & 0.09 & 0.15 & -0.044 \\
NGC 4472 & -21.81 & 3.93 & 0.43 & 0.01 & -0.040 \\
NGC 4552 & -20.44 & 3.36 & 0.28 & -0.10 & -0.080 \\
NGC 4649 & -21.51 & 3.80 & 0.42 & 0.16 & -0.053 \\
NGC 4697 & -20.36 & 3.65 & 0.71 & 0.22 & --- \\
NGC 5638 & -20.12 & 3.65 & 0.73 & --- & --- \\
NGC 5812 & -20.40 & 3.47 & 0.52 & 0.59 & --- \\
NGC 5813 & -21.20 & 3.90 & 0.51 & -0.10 & -0.045 \\
NGC 5831 & -19.94 & 3.57 & 0.19 & 0.33 & -0.078 \\
NGC 6703 & -20.21 & 3.50 & 0.30 & --- & --- \\
\hline 
\multicolumn{6}{c}{Other galaxies from G93:}\\
NGC 0221 & -15.91 & 2.20 & 0.89 & 0.00 & --- \\
NGC 0315 & -22.15 & 4.23 & 0.09 & --- & --- \\
NGC 0507 & -21.74 & 4.35 & 0.09 & 0.00 & --- \\
NGC 0547 & -21.20 & 3.91 & 0.24 & --- & --- \\
NGC 0636 & -20.23 & 3.45 & 1.04 & --- & --- \\
NGC 1453 & -21.33 & 3.85 & 0.62 & --- & --- \\
NGC 1600 & -22.23 & 4.17 & 0.03 & -0.03 & --- \\
NGC 2300 & -20.38 & 3.65 & 0.08 & 0.07 & --- \\
NGC 2778 & -18.67 & 3.34 & 0.74 & 0.33 & --- \\
NGC 3377 & -19.26 & 3.28 & 0.86 & 0.03 & -0.075 \\
NGC 3379 & -20.02 & 3.27 & 0.72 & 0.18 & -0.043 \\
NGC 3818 & -20.41 & 3.58 & 0.93 & --- & --- \\
NGC 4261 & -21.22 & 3.79 & 0.10 & 0.16 & -0.089 \\
NGC 4478 & -19.16 & 3.11 & 0.84 & -0.10 & -0.103 \\
NGC 4489 & -18.46 & 3.46 & 1.49 & --- & --- \\
NGC 6127 & -21.03 & 3.82 & 0.11 & --- & --- \\
NGC 6702 & -20.55 & 3.87 & 0.18 & --- & --- \\
NGC 7052 & -21.14 & 3.96 & 0.34 & 0.16 & --- \\
NGC 7454 & -19.34 & 3.49 & 0.13 & --- & --- \\
NGC 7562 & -21.50 & 3.86 & 0.06 & --- & --- \\
NGC 7619 & -21.77 & 3.93 & 0.53 & -0.02 & --- \\
NGC 7626 & -21.03 & 3.88 & 0.12 & 0.47 & --- \\
NGC 7785 & -20.91 & 3.78 & 0.47 & -0.10 & --- \\
\enddata
\label{supp}
\end{deluxetable}

\begin{deluxetable}{lcl}
\tablecaption{Principal Component Analysis Table} 
\tablecolumns{3}
\tablewidth{0pt}
\tablehead{
\colhead{Input Parameters} & \colhead{\% of variance} & \colhead{Principal Component} \\
}
\startdata
M$_V$, $\gamma$, log($t$) & 52 & ${\rm PC1} = -0.305{\rm M}_V - 0.681\gamma + 0.666{\rm log}(t)$ \\
\hline
M$_V$, $\gamma$, [Z/H] & 54 & ${\rm PC1} = 0.170{\rm M}_V + 0.731\gamma + 0.661{\rm [Z/H]}$ \\
\hline
M$_V$, $\gamma$, [$\alpha$/Fe] & 45 & ${\rm PC1} = -0.258{\rm M}_V - 0.702\gamma + 0.663[\alpha/{\rm Fe}]$ \\
\\
 & 38 & ${\rm PC2} = 0.665\gamma + 0.741[\alpha/{\rm Fe}]$ \\
\hline
M$_V$, log($r_b$), log($t$) & 64 & ${\rm PC1} = 0.634{\rm M}_V - 0.618{\rm log}(r_b) - 0.466{\rm log}(t)$ \\
\hline
M$_V$, log($r_b$), [Z/H] & 49 & ${\rm PC1} = 0.433{\rm M}_V - 0.898{\rm log}(r_b)$ \\
\\
 & 39 & ${\rm PC2} = 0.111{\rm M}_V + 0.993{\rm [Z/H]}$ \\
\hline
M$_V$, log($r_b$), [$\alpha$/Fe] & 56 & ${\rm PC1} = 0.328{\rm M}_V - 0.735{\rm log}(r_b) - 0.594[\alpha/{\rm Fe}]$ \\
\hline
M$_V$, $\gamma$, log($r_b$), log($\sigma$), & 38 & ${\rm PC1} = -0.275{\rm M}_V - 0.346\gamma + 0.460{\rm log}(r_b) +$ \\
log($t$), [Z/H], [$\alpha$/Fe] & & $+ 0.467{\rm log}(\sigma) + 0.446{\rm log}(t) - 0.200{\rm [Z/H]} +$ \\
& & $+ 0.369[\alpha/{\rm Fe}]$ \\
\\
 & 25 & ${\rm PC2} = 0.287\gamma + 0.160{\rm log}(r_b) + 0.373{\rm log}(\sigma) -$ \\
 & & $- 0.360{\rm log}(t) + 0.702{\rm [Z/H]} + 0.351[\alpha/{\rm Fe}]$ \\
\\
 & 20 & ${\rm PC3} = -0.591\gamma + 0.527{\rm log}(r_b) - 0.304{\rm log}(\sigma) -$ \\
 & & $- 0.436{\rm log}(t) + 0.175{\rm [Z/H]} - 0.242[\alpha/{\rm Fe}]$ \\  
\enddata
\label{pca}
\end{deluxetable}
\clearpage

\vbox{
\begin{center}
\includegraphics[width=\textwidth]{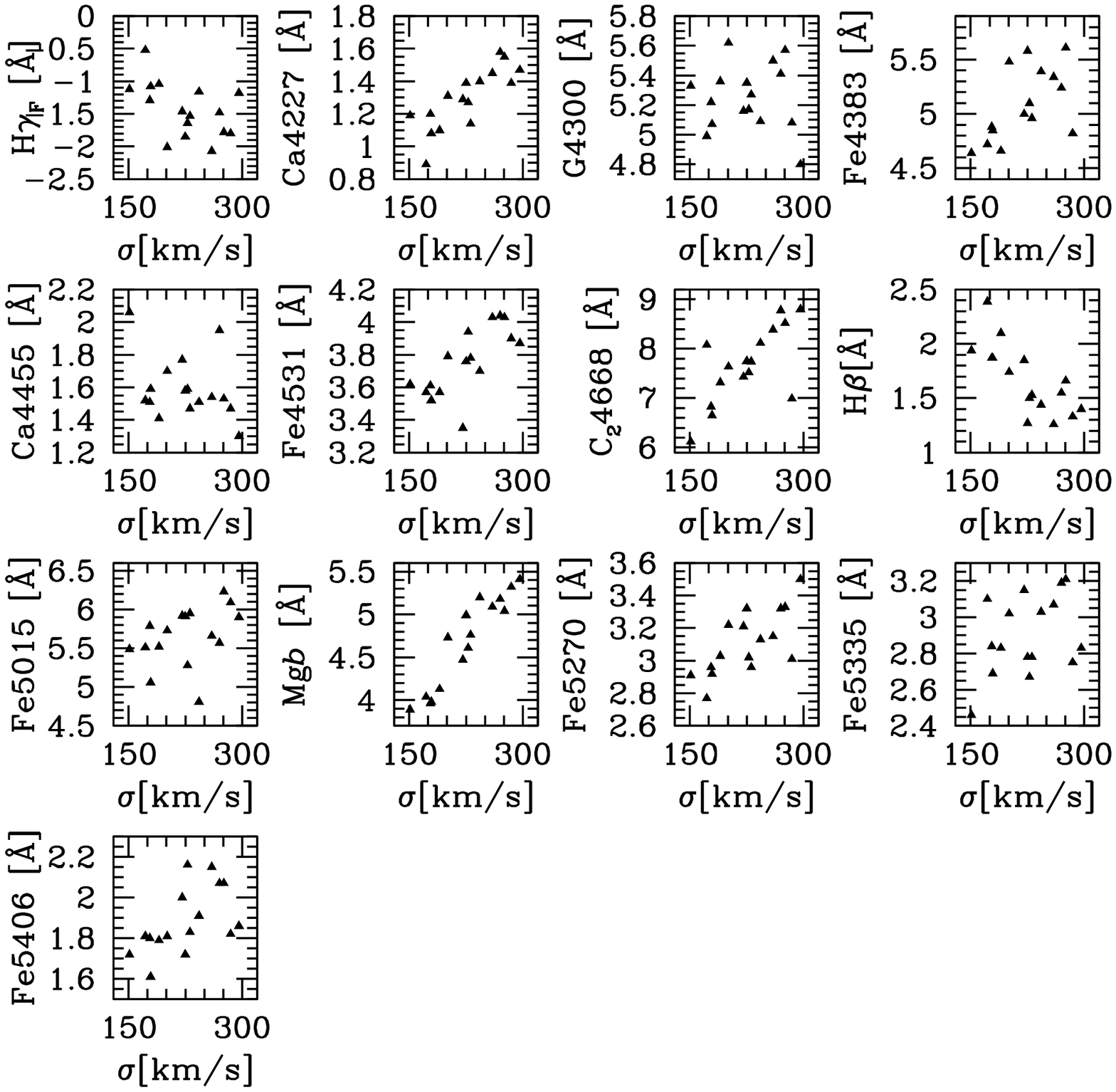}
\figcaption{\small
Spectral indices plotted against galaxy velocity dispersion.  
The Mg--$\sigma$ relation is clearly apparent, as are similar but noiser
relations in Ca4227, ${\rm C}_24668$, and the iron indices.  A relation
in the opposite direction (smaller index value at larger $\sigma$) is seen
in H$\beta$ and H$\gamma$.
\label{index}
}
\end{center}}

\vbox{
\begin{center}
\includegraphics[width=\textwidth]{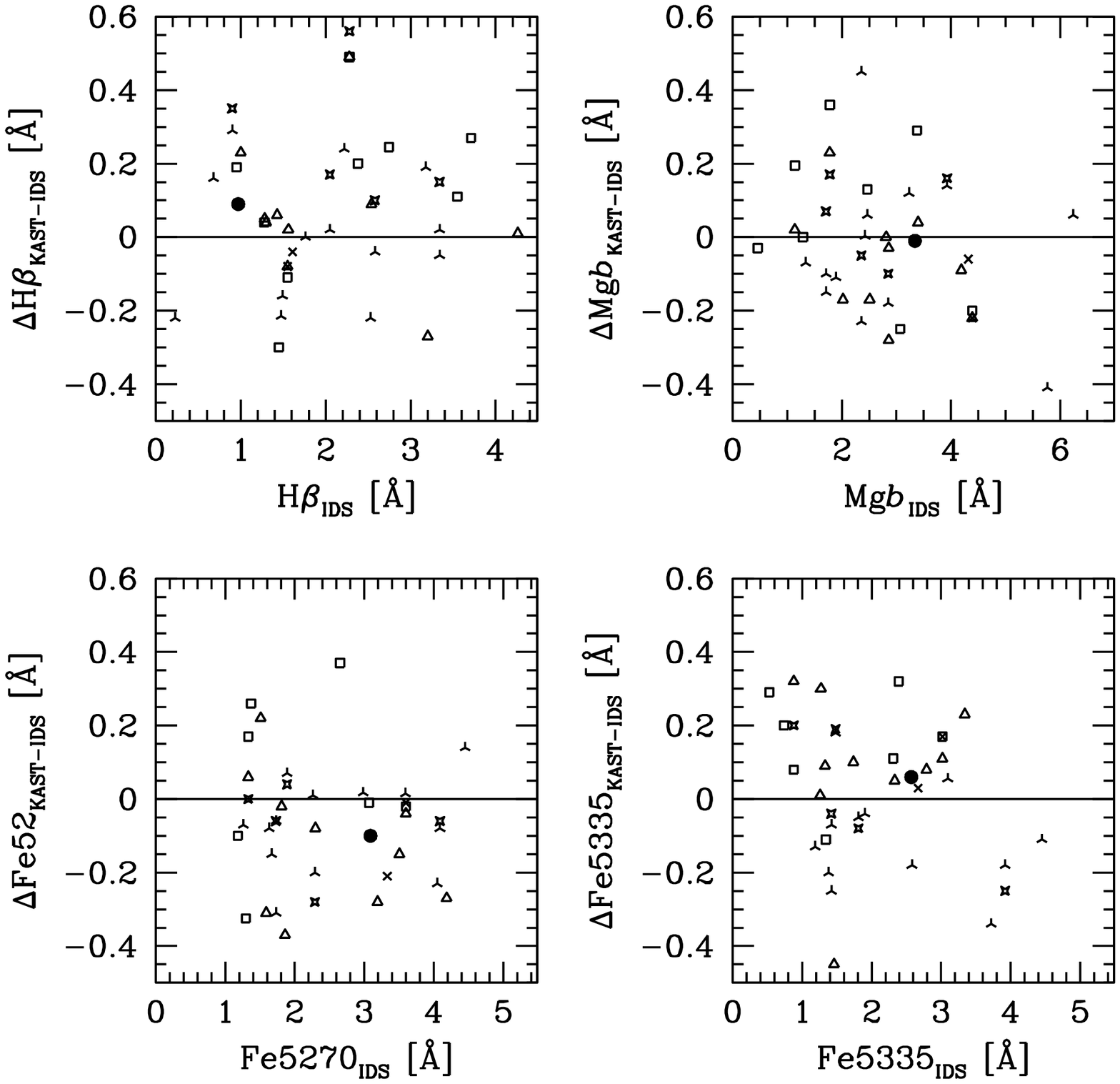}
\figcaption{\small
Comparisons of standard star primary index measurements to the Lick/IDS
sample of \citet{ch1worthey94}.  Different point types represent different
observing runs.  The large solid circle represents HD~51440, an internal 
standard star for the Lick/IDS system and as such the most accurately 
measured data point in this figure.  No calibration correction has been 
applied to these data points.  The calibration uncertainty shown here is
a major limitation on the overall index uncertainty, as the calibration
error often dominates the error budget.
\label{stdpre}
}
\end{center}}

\vbox{
\begin{center}
\includegraphics[width=\textwidth]{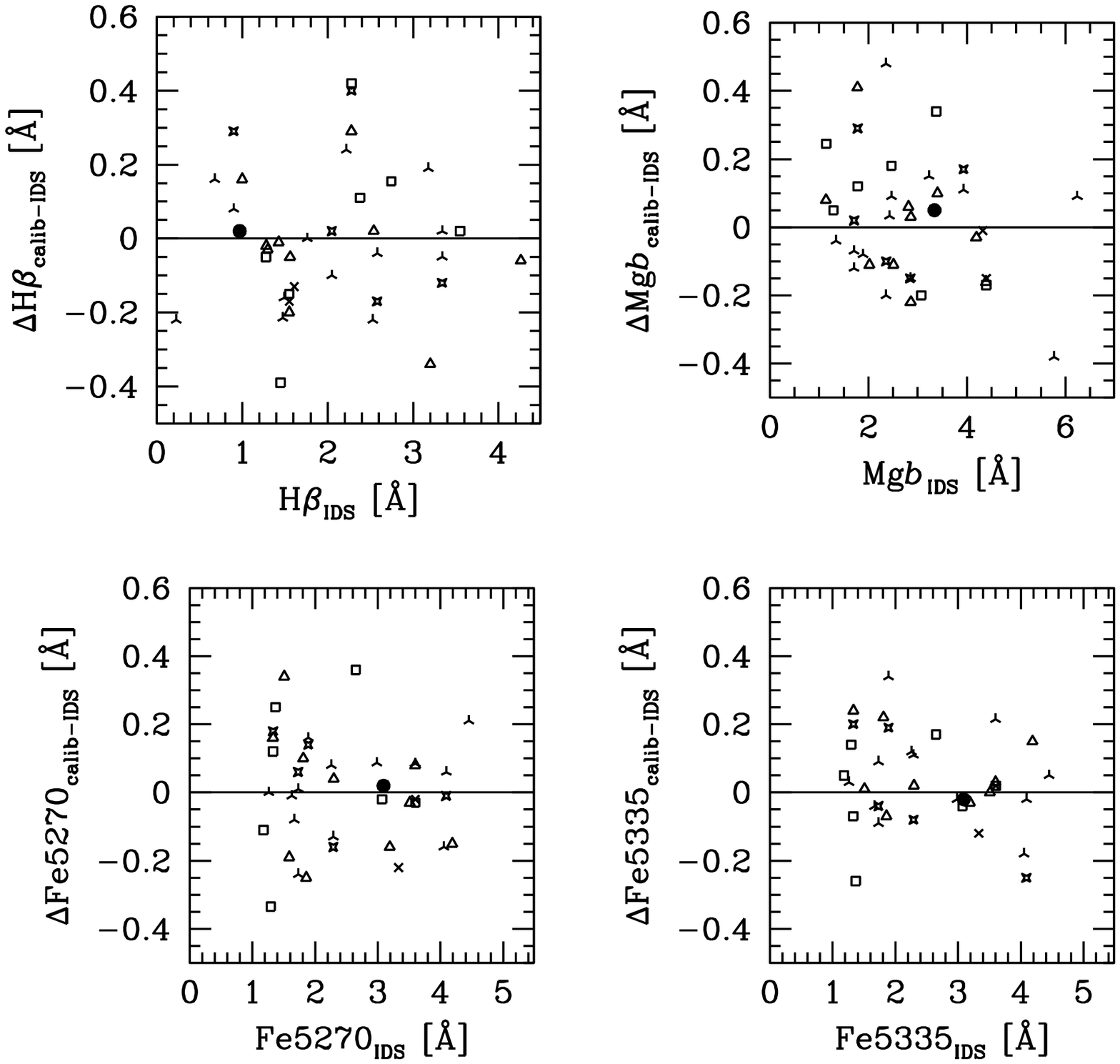}
\figcaption{\small
Comparisons of standard star primary index measurements to the Lick/IDS
sample of \citet{ch1worthey94}.  Different point types represent different
observing runs as above.  A constant calibration offset has been applied 
to each observing run; the 2002 November and 2003 January runs have been 
combined for this purpose due to the limited number of standard stars 
observed in each.  No trend with index strength is seen, confirming that 
a constant offset is sufficient to calibrate the data onto the Lick/IDS system.
\label{stdpost}
}
\end{center}}

\vbox{
\begin{center}
\includegraphics[width=\textwidth]{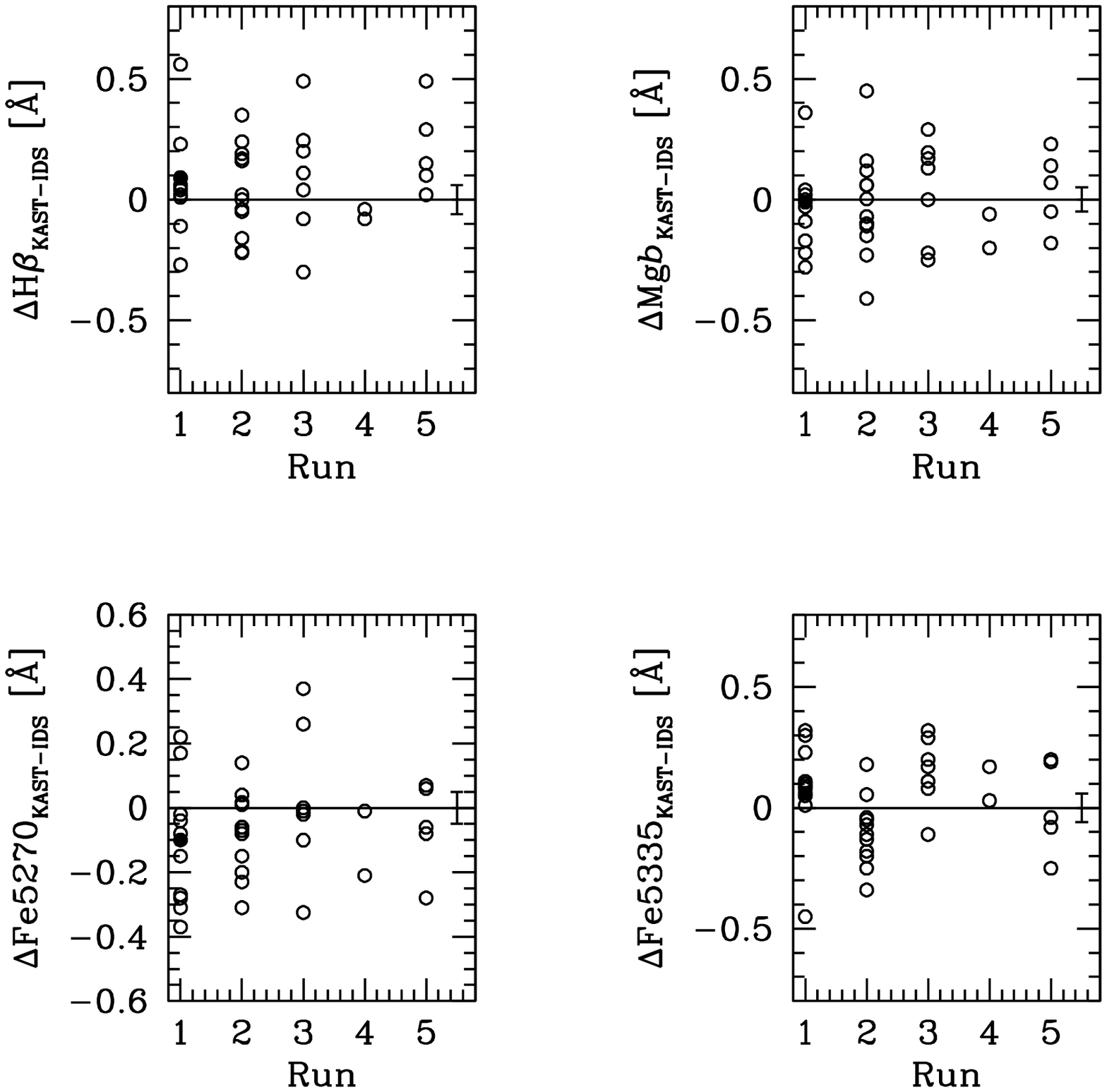}
\figcaption{\small
Comparisons of uncorrected standard star primary index measurements to the 
Lick/IDS sample of \citet{ch1worthey94}.  The solid point is HD~51440, one
of the internal standard stars for the IDS system; as such it is more
accurately measured than any other star in this data set.  Also shown is
the typical error in the mean for the index offsets within a given 
observing run.  The run to run variations in the mean index values are
apparent.
\label{stdtpre}
}
\end{center}}

\vbox{
\begin{center}
\includegraphics[width=\textwidth]{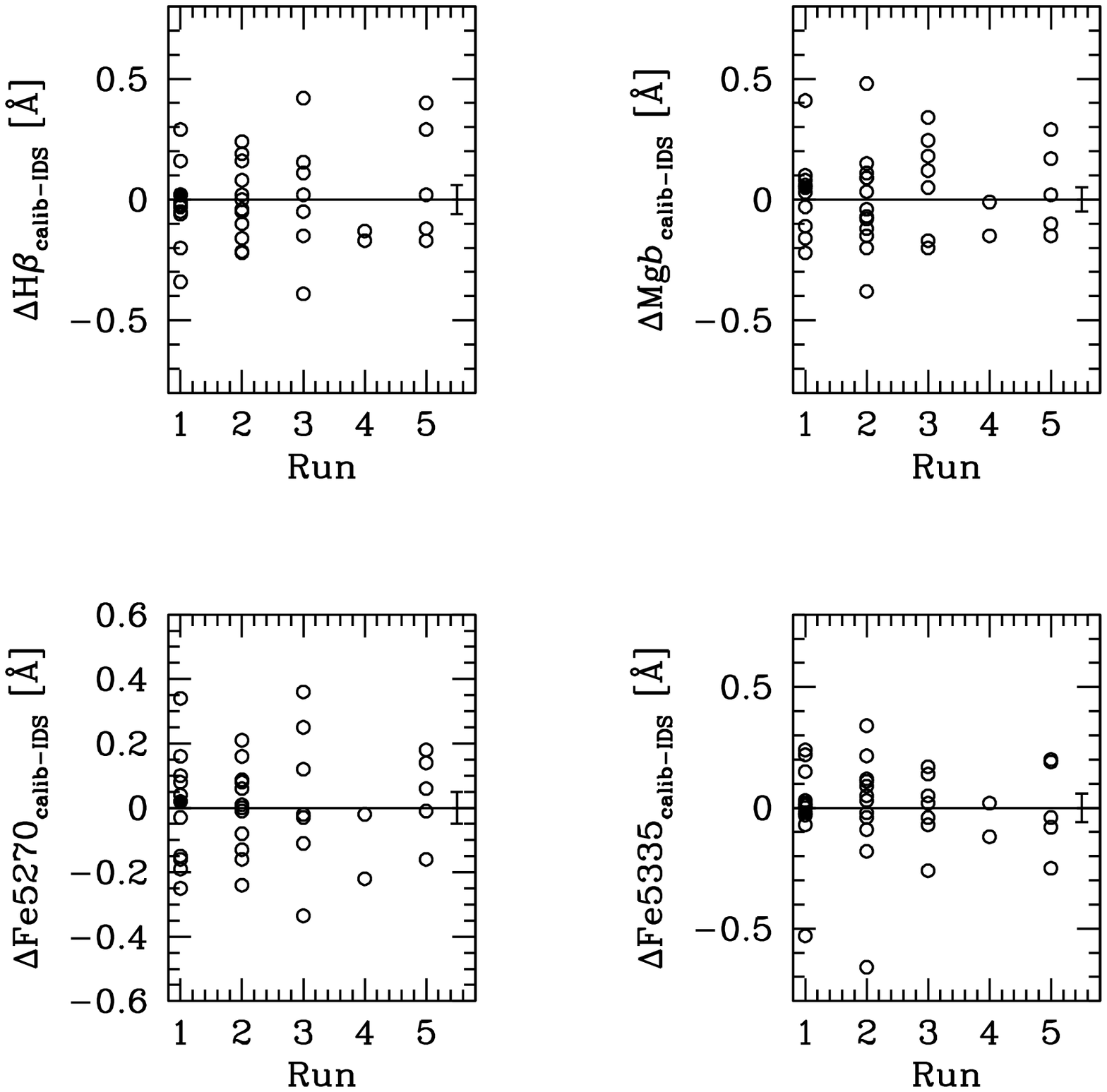}
\figcaption{\small
As in Fig.~\ref{stdtpre}, after correcting the observations to the Lick/IDS
system.  The offset for each run was determined separately except for
Runs~4 and 5 which were combined for calibration purposes.
\label{stdtpost}
}
\end{center}}

\vbox{
\begin{center}
\includegraphics[width=\textwidth]{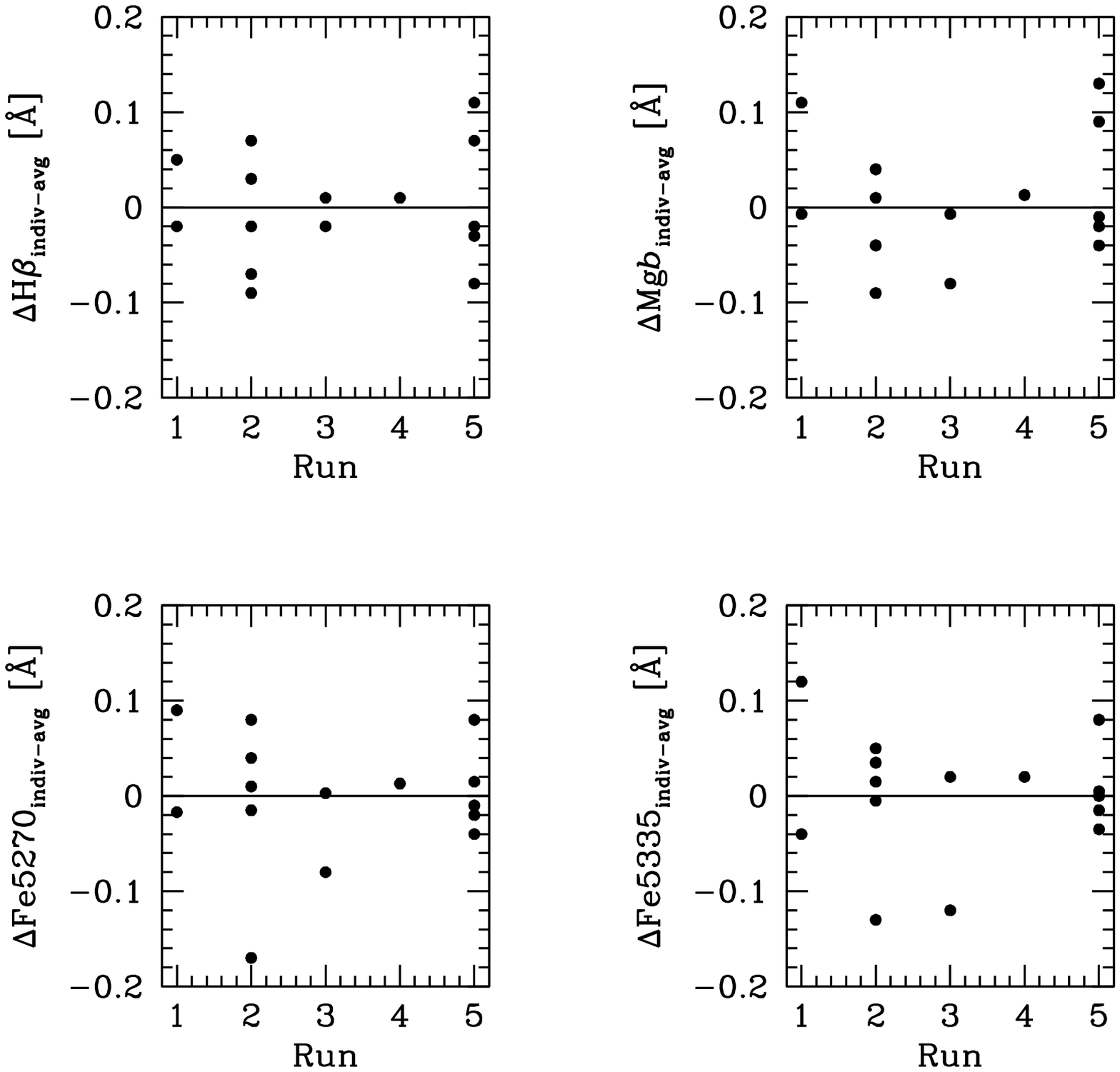}
\figcaption{\small
Stars which were observed repeatedly during the observing program are
shown.  Unlike in Figs.~\ref{stdtpre} and \ref{stdtpost}, individual index
measurements are compared to the average index measurement for each star,
not to the Lick/IDS system.  Thus this shows the internal error in each
index, independent of the uncertainty in calibrating to the Lick/IDS system.
\label{stdint}
}
\end{center}}

\vbox{
\begin{center}
\includegraphics[width=\textwidth]{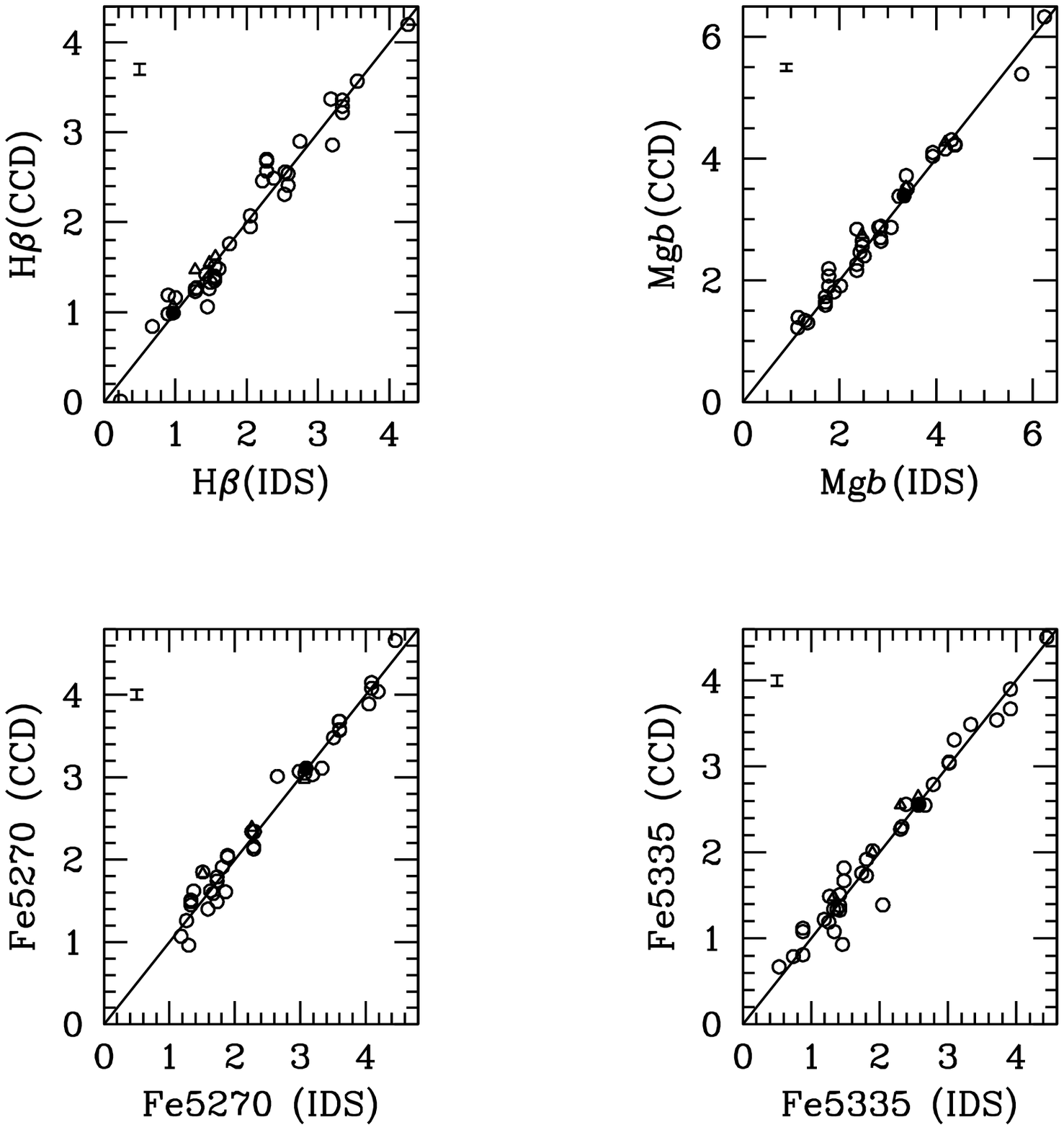}
\figcaption{\small
The calibrated primary index measurements for all standard stars are
plotted against the standard value for that star on the Lick/IDS system.
Open circles are ordinary standard stars from this work, and the solid
circle as above is the Lick/IDS internal standard star HD~51440.  
Several stars were observed as part of both G93 and this study; the 
calibrated G93 measurements for those stars are shown as open triangles.  
The adopted calibration is in good agreement with both the Lick/IDS system 
and the G93 calibration.
\label{stdchk}
}
\end{center}}

\vbox{
\begin{center}
\includegraphics[width=\textwidth]{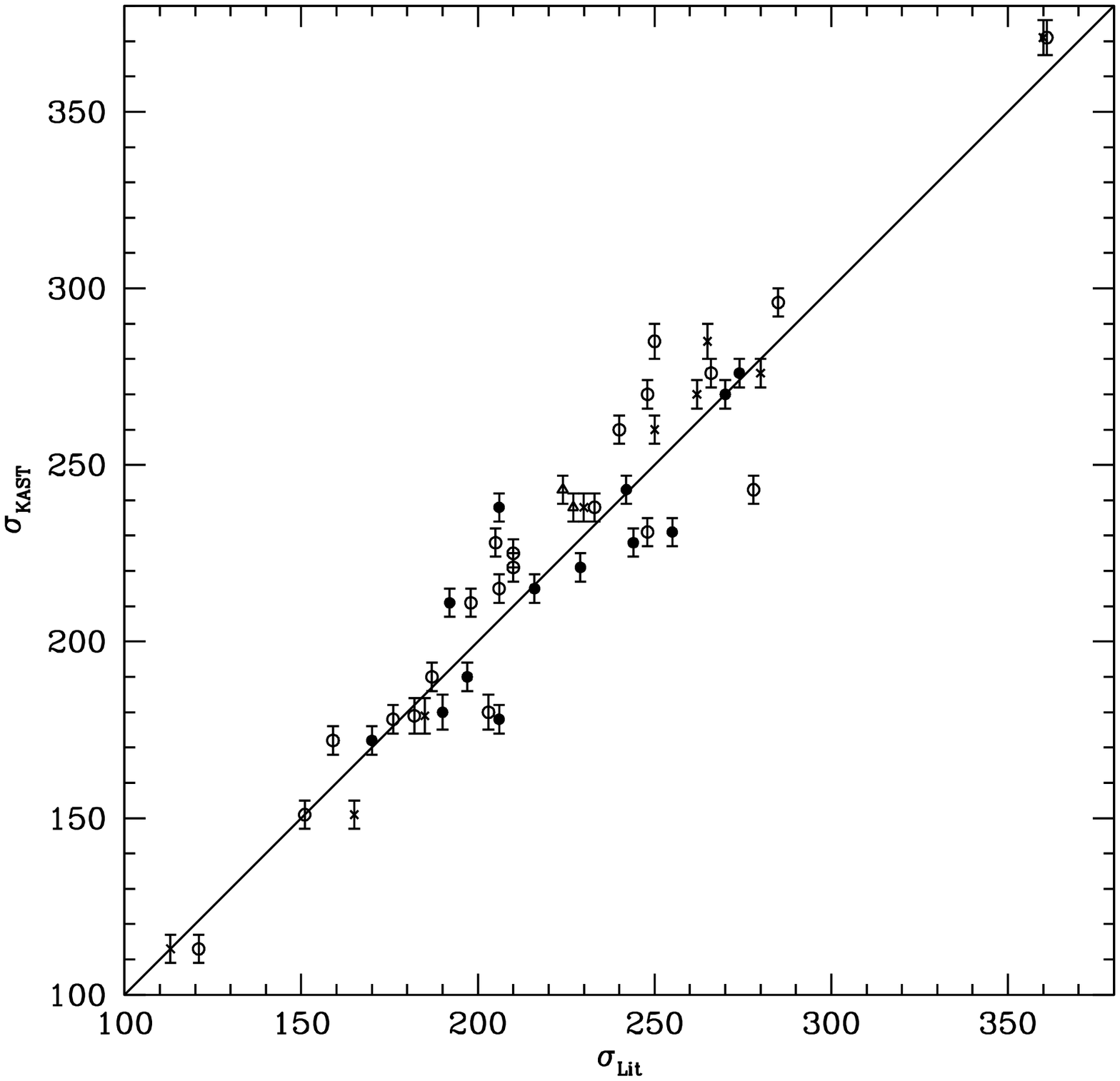}
\figcaption{\small
Comparisons of measured velocity dispersions between galaxies in this study 
and in other data sets.  Open triangles are compared with G93. Black open 
circles, x's, and solid circles are compared with \citet{ch1faber89}, 
\citet{faber97}, and \citet{ch1denicolo} respectively.  The velocity 
dispersion measurements from this study are in good agreement with previous work.
\label{vdcomp}
}
\end{center}}

\vbox{
\begin{center}
\includegraphics[width=\textwidth]{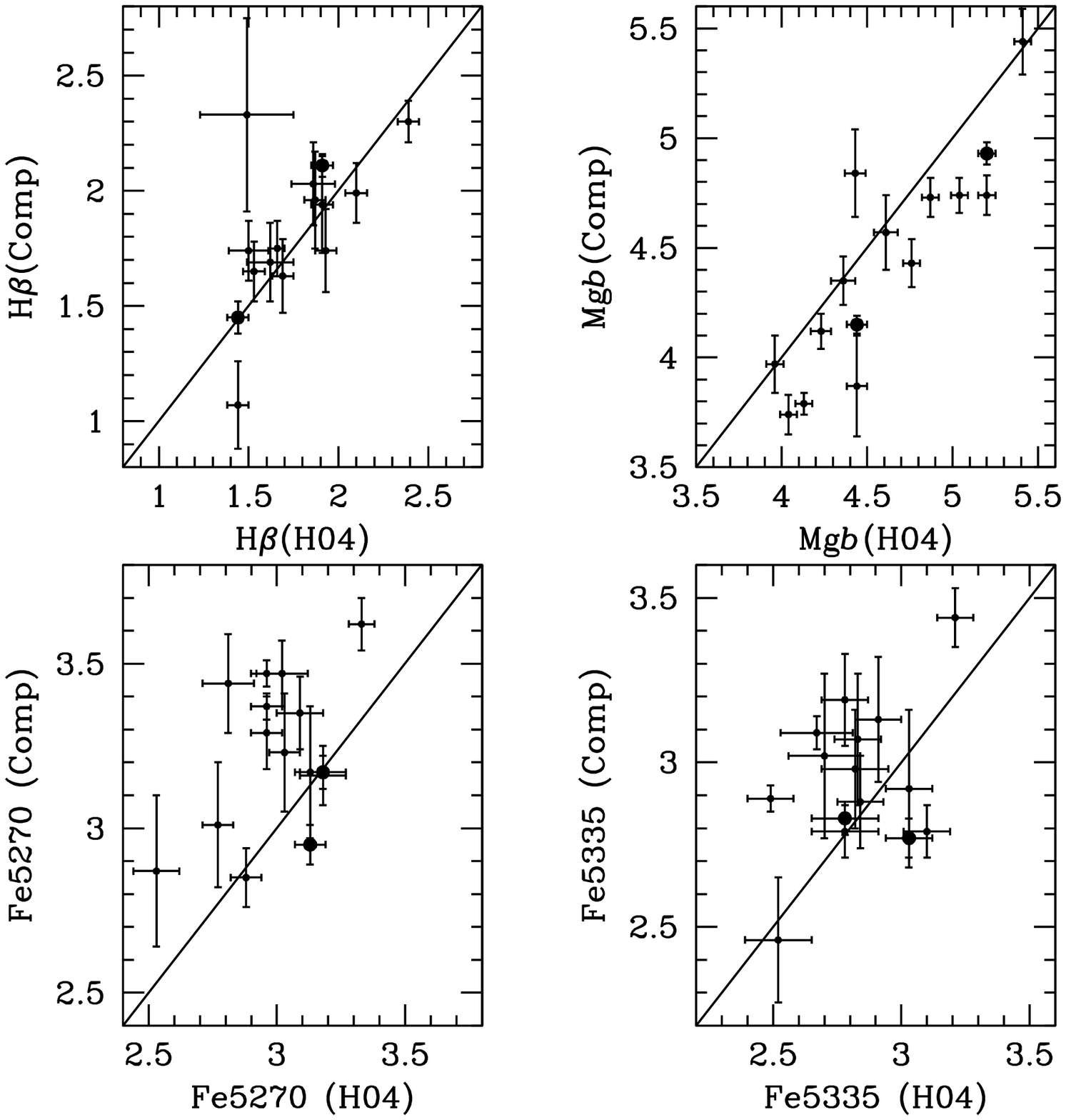}
\figcaption{\small
Comparisons of measured spectral indices between galaxies in this study and 
in other data sets.  Points in large type are compared with G93.  Points in 
small type are compared with \citet{ch1denicolo}.
\label{comp}
}
\end{center}}

\vbox{
\begin{center}
\includegraphics[width=\textwidth]{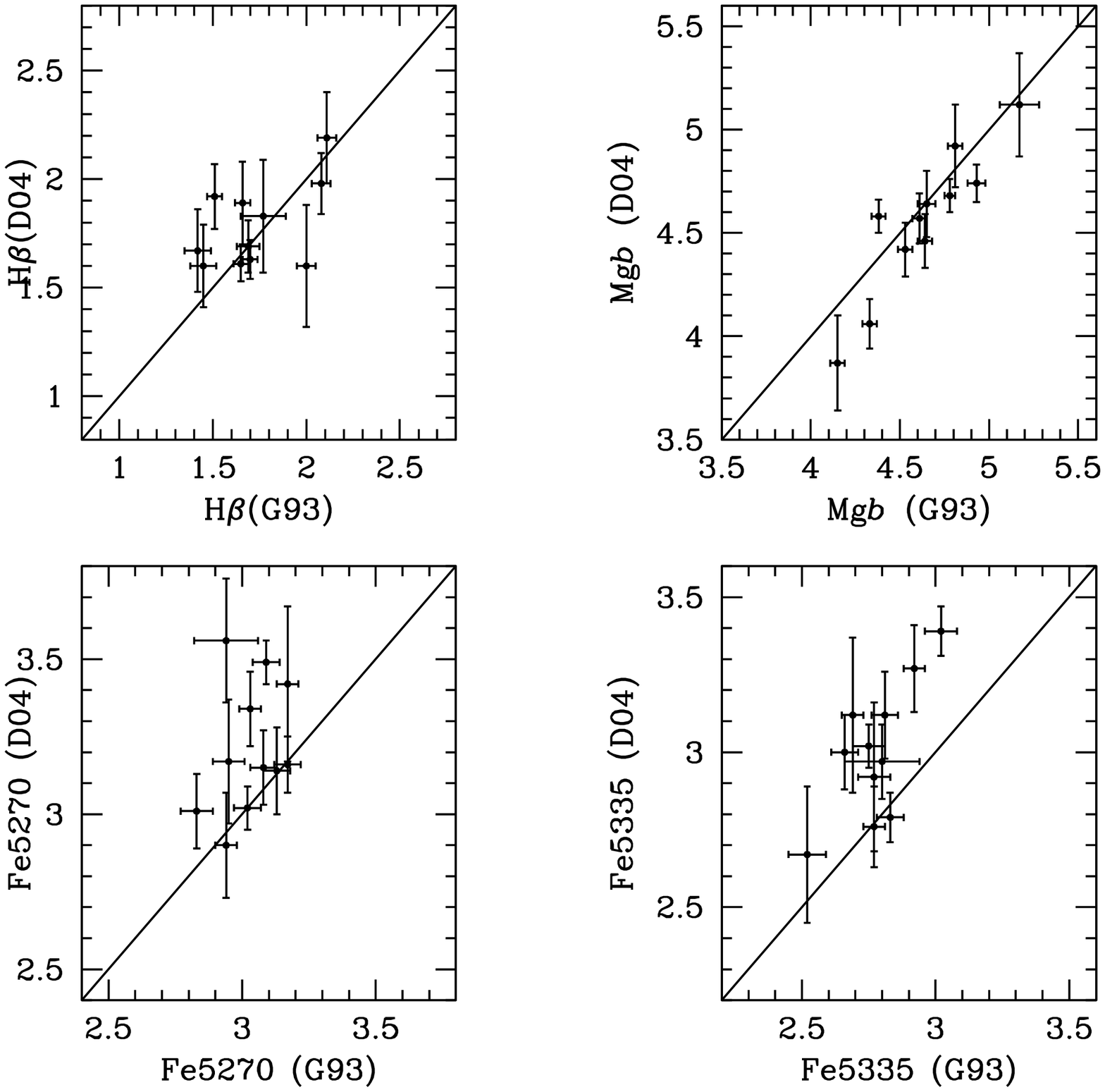}
\figcaption{\small
Comparisons of measured velocity dispersion and spectral indices between 
galaxies in G93 and in \citet{ch1denicolo}.  The fact that the 
\citet{ch1denicolo} data set deviates from the G93 in each quantity in the 
same way that the \citet{ch1denicolo} data set deviates from the volume-limited 
data set (Fig.~\ref{comp}) lends confidence that the volume-limited data set 
has been successfully calibrated onto the Lick/IDS system.
\label{gdcomp}
}
\end{center}}

\vbox{
\begin{center}
\includegraphics[width=\textwidth]{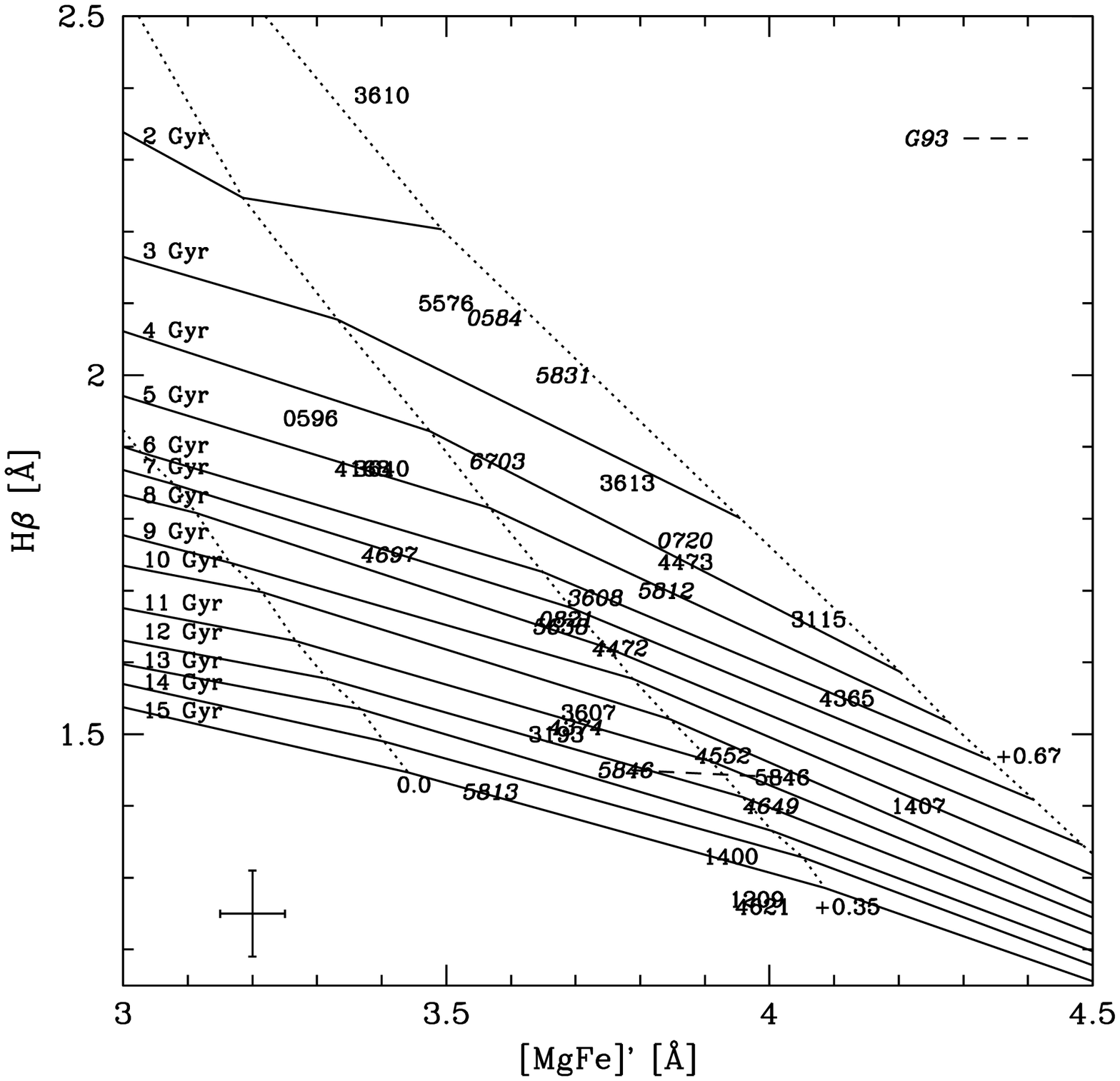}
\figcaption{\small
The galaxies in the volume-limited sample are plotted on age-metallicity
grids from TMB.  The horizontal axis is the [MgFe]$^\prime$ index defined by 
TMB to be independent of $[\alpha/{\rm Fe}]$.  Galaxies in normal type are from 
this study, while galaxies in italics are members of
the volume-limited sample from the G93 data set.  A dashed line connects
the measurements of NGC~5846 from the two data sets; this galaxy provides a
direct empirical test of the accuracy of the index measurements and consistency
of the Lick/IDS calibration.  Although the H$\beta$ measurements are in 
excellent agreement between the two studies, the Mg$b$, Fe5270, and Fe5335
index measurements are discrepant by more than two standard deviations.
Much of this discrepancy results from the unusually large difference in 
measured velocity dispersion between the two studies.
Typical error bars are shown in the lower left.  The volume-limited sample
is in agreement with previous work, showing a wide range of ages 
and a smaller range of supersolar metallicities.
\label{grid}
}
\end{center}}

\vbox{
\begin{center}
\includegraphics[width=\textwidth]{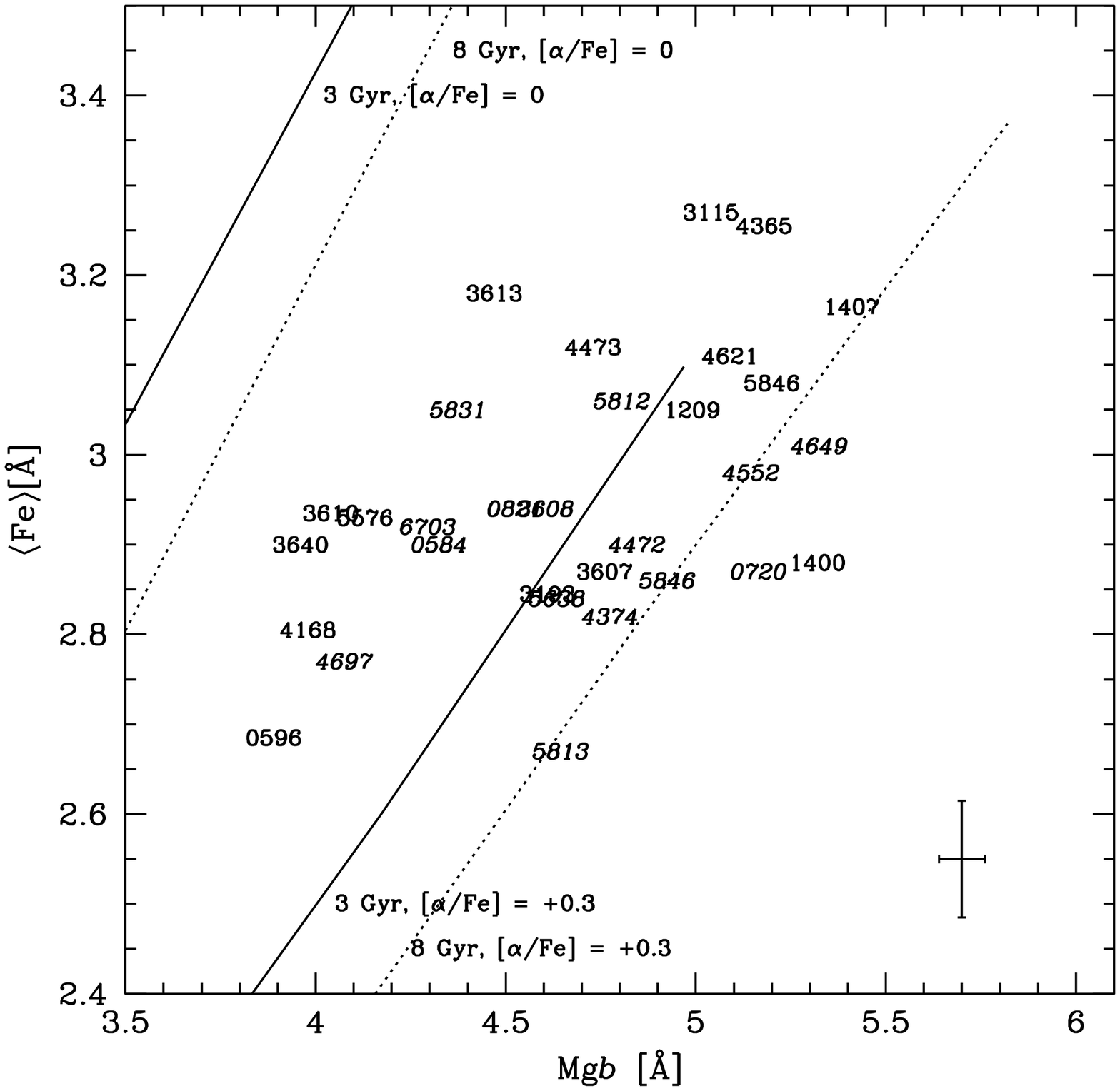}
\figcaption{\small
The galaxies in the volume--limited sample are plotted against iron and
magnesium lines to measure $[\alpha/{\rm Fe}]$.  Models from
TMB are shown, at 3~Gyr age (solid lines) and 8~Gyr age (dotted lines),
and $[\alpha/{\rm Fe}]$ values of 0.0 and +0.3 dex.  Galaxies are presented 
as in Fig.~\ref{grid}.  Typical error bars are shown in the lower right
corner.  The two sets of models shown illustrate the relatively
small effect of age on $[\alpha/{\rm Fe}]$ measurements.
\label{alpha}
}
\end{center}}

\vbox{
\begin{center}
\includegraphics[width=\textwidth]{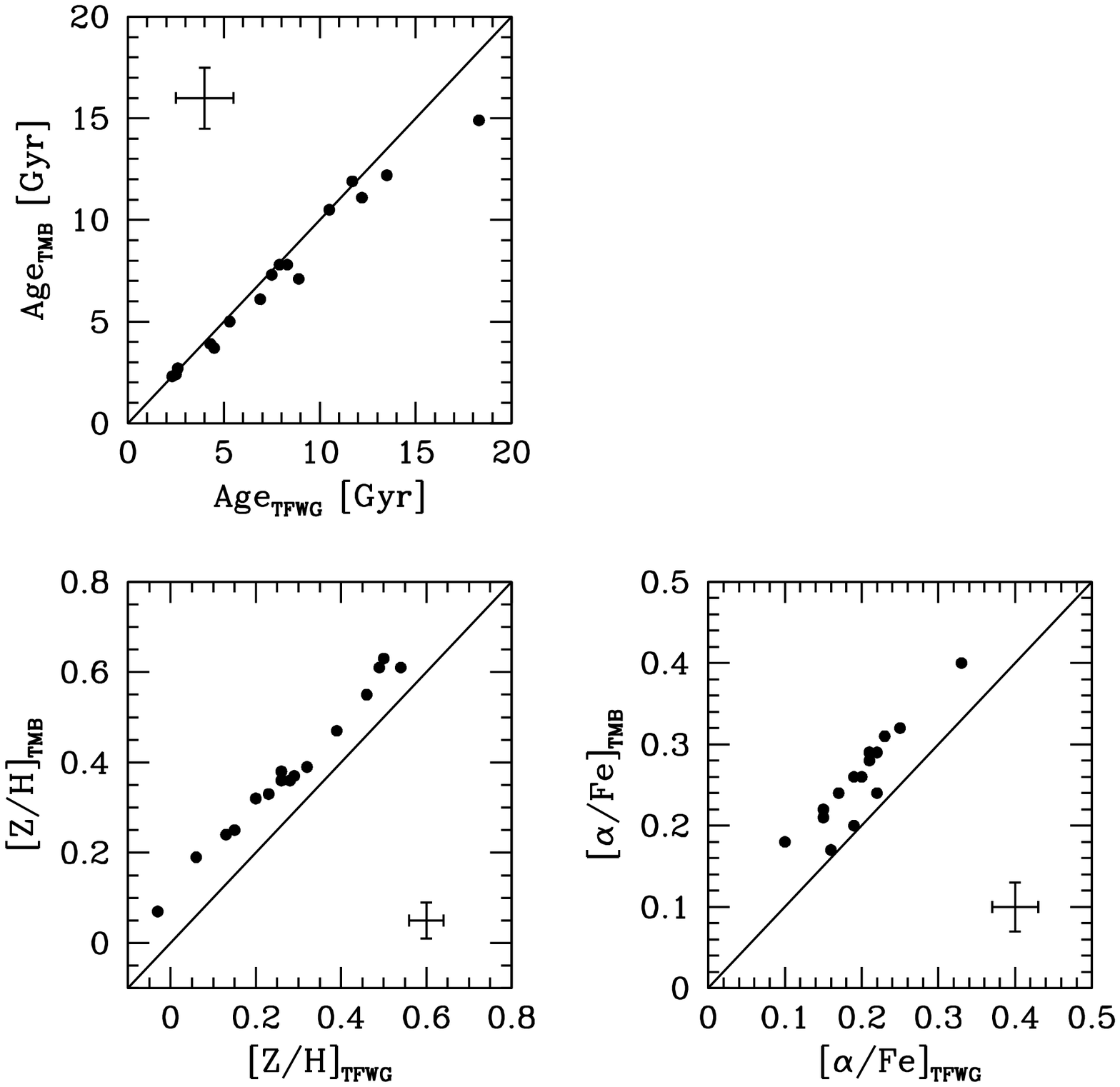}
\figcaption{\small
Derived physical parameters for the G93 sample are compared.  Parameters
from the TMB models are on the vertical axes, while parameters from 
\citet{ch1tfwg1} are on the horizontal axes.  Age is compared in the upper
left, metallicity in the lower left, and $[\alpha/{\rm Fe}]$ in the
lower right.  Representative error bars are shown.
The three galaxies with similar $[\alpha/{\rm Fe}]$ using
both models are NGC~1700, NGC~5831, and NGC~584, the three youngest galaxies
plotted.  The qualitative difference between the two models is negligible.
TMB models result in a constant offset toward larger $[\alpha/{\rm Fe}]$, 
which in turn results in larger [Z/H].
\label{diff}
}
\end{center}}

\vbox{
\begin{center}
\includegraphics[width=\textwidth]{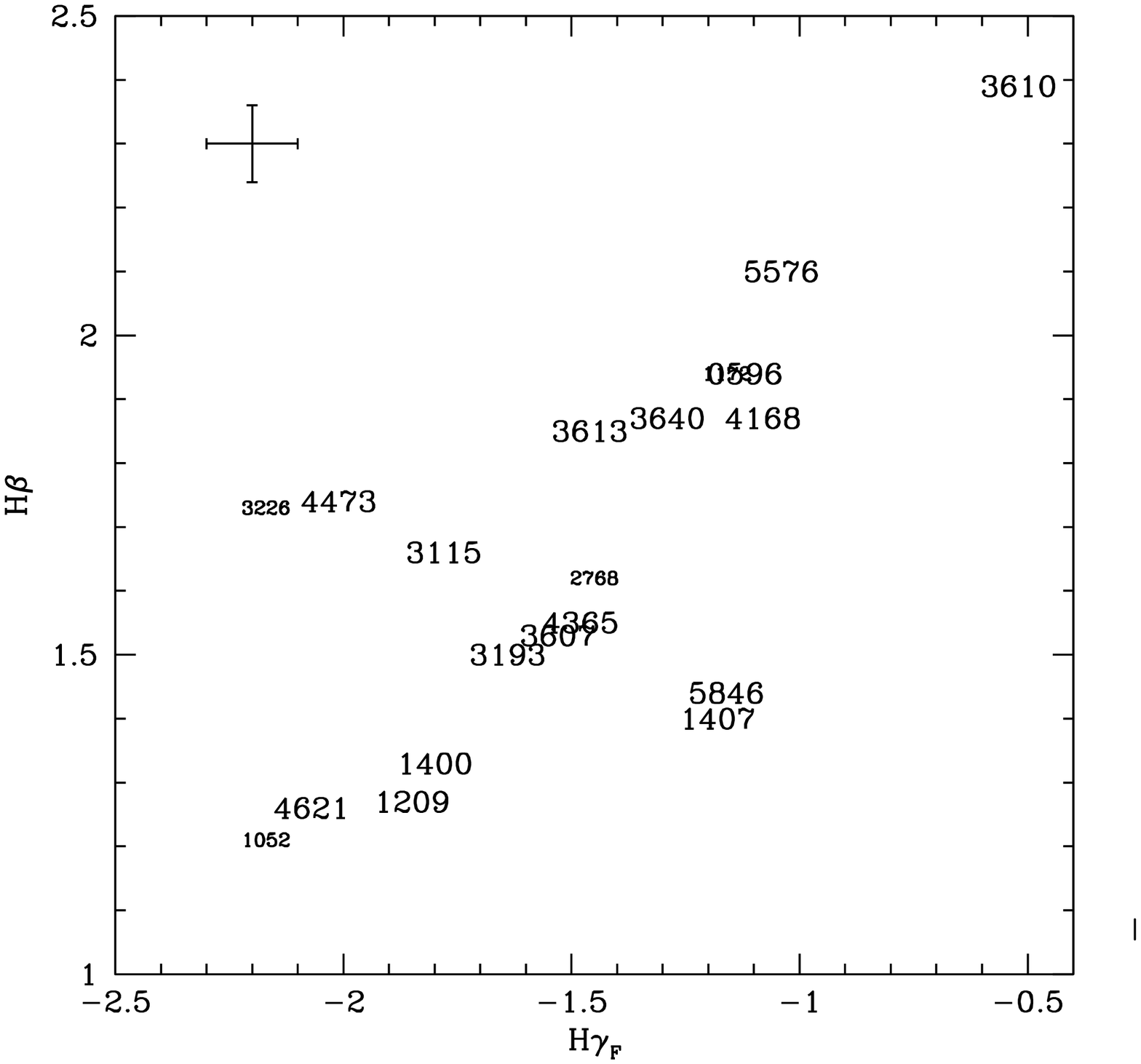}
\figcaption{\small
Comparison of H$\beta$ and H$\gamma_{\rm F}$ index measurements.  The
galaxies in small type have large emission corrections and correspondingly
large uncertainties.  Typical uncertainties for the rest of the galaxy
sample are presented in the upper left.  With the exception of four 
outliers, the Balmer index measurements follow a tight linear relation.
\label{balmercomp}
}
\end{center}}

\vbox{
\begin{center}
\includegraphics[width=\textwidth]{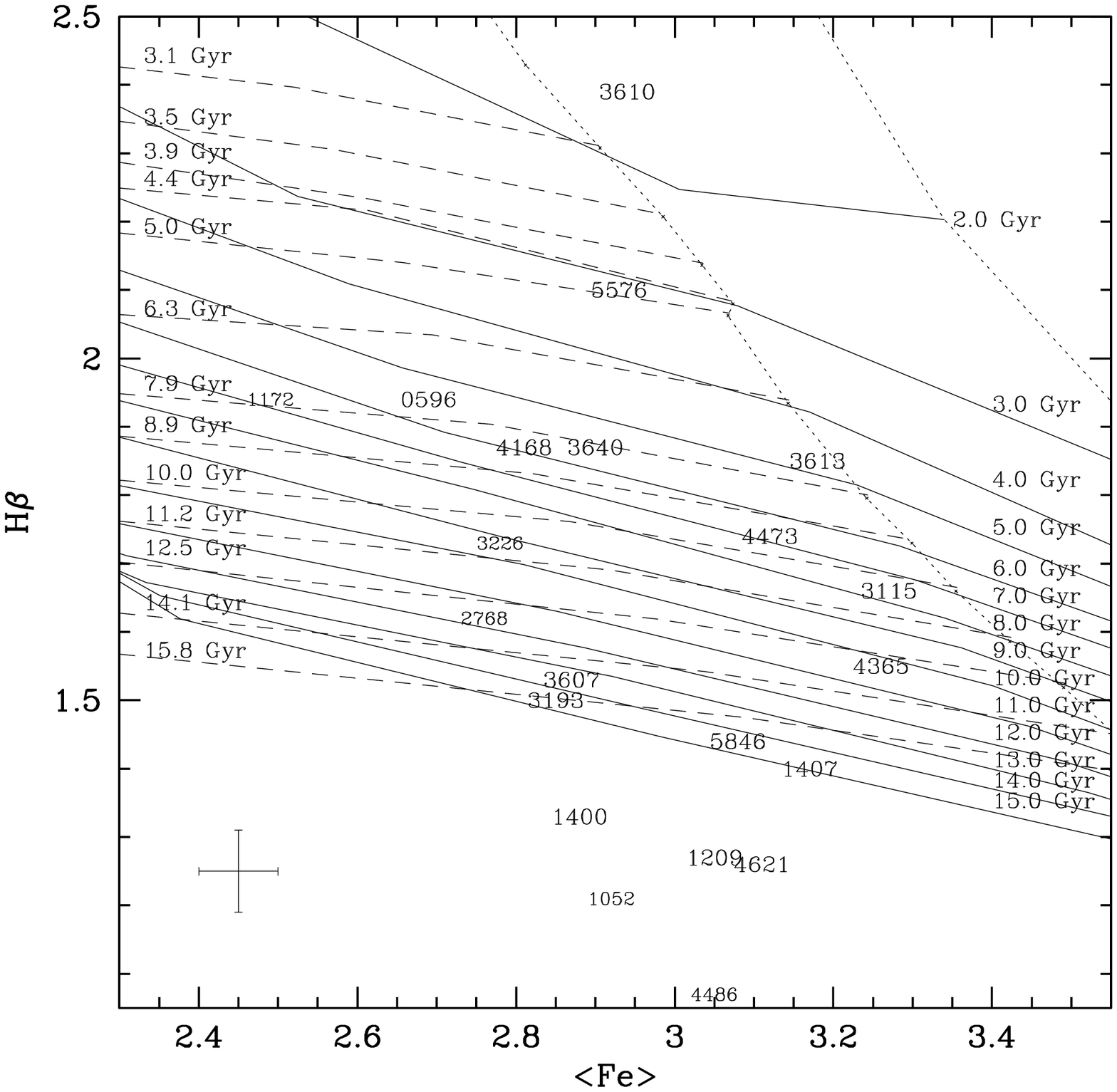}
\figcaption{\small
The galaxy sample from this work is plotted on the H$\beta$ vs. 
$\langle{\rm Fe}\rangle$ plane, along with two sets of model grids.  Models 
in solid type are from TMB with [$\alpha$/Fe]$=0.0$,
while models in dashed type are from \citet{ch1schiavon}.  In both sets of 
models only the most metal-rich isometallicity line (TMB: [Z/H]$=+0.67$; 
Schiavon: [Fe/H]$=+0.2$) is plotted (dotted lines).  Isochrone lines for 
each set of models are labeled.  As above, galaxies in small type
have large emission corrections.  Typical uncertainties for the remaining
galaxies are presented in the lower left.  A consistent age offset is
apparent between the two sets of models, in the sense that the Schiavon
models measure older ages by $\sim2$~Gyr compared to the TMB models.
\label{hbfe}
}
\end{center}}

\vbox{
\begin{center}
\includegraphics[width=\textwidth]{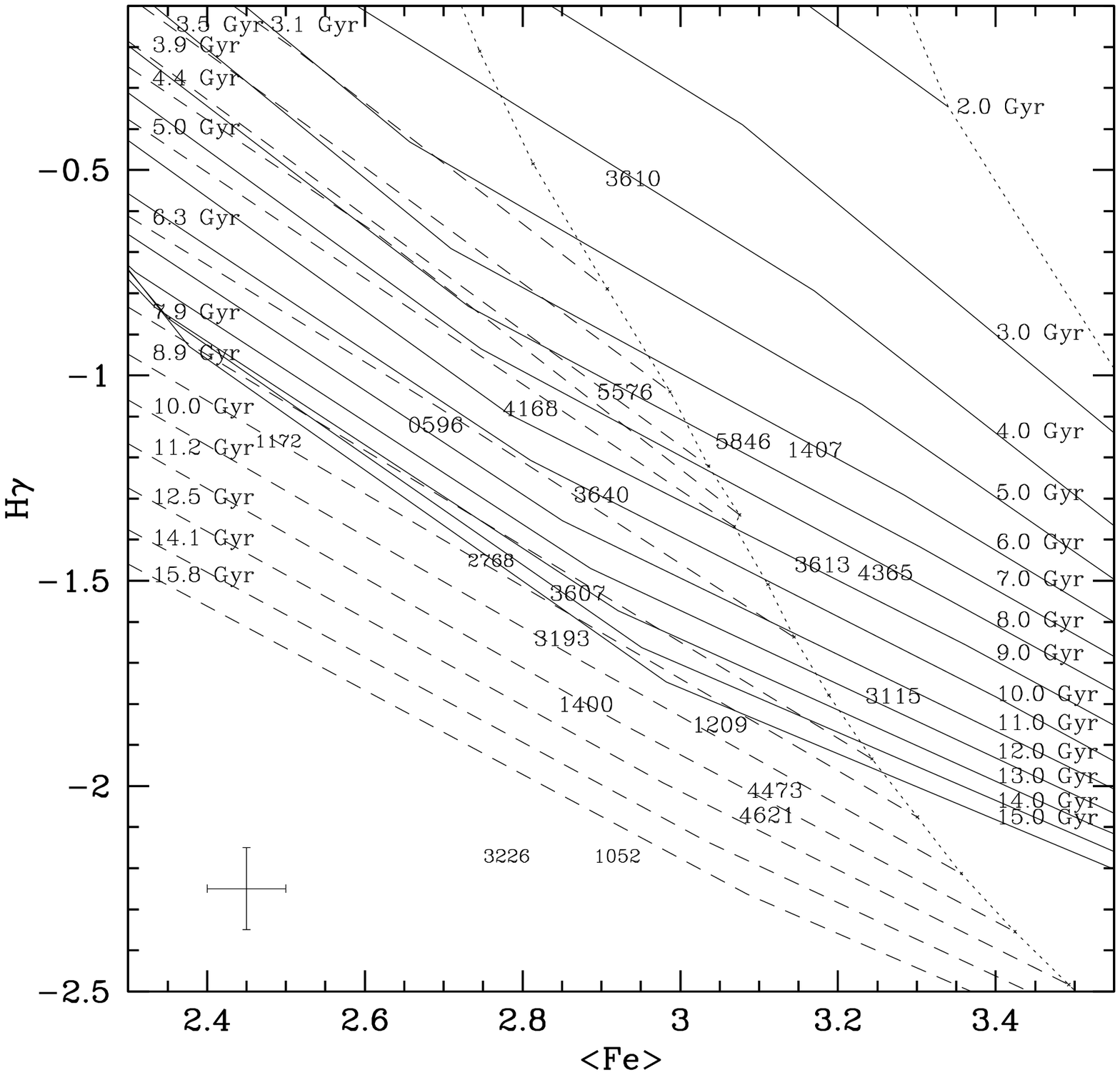}
\figcaption{\small
Galaxies and dashed models are as in Fig.~\ref{hbfe}.  Models in solid type
are from \citet*{ch1tmk04} with [$\alpha$/Fe]$=0.0$.  The two sets of models 
are substantially different from each other in the H$\gamma_{\rm F}$ vs. 
$\langle{\rm Fe}\rangle$ plane.  This is primarily due to the \citet{ch1tmk04} 
models predicting larger H$\gamma_{\rm F}$ equivalent widths by $\sim0.4$~\AA~
compared to the Schiavon models.  As a result, age measurements using
the Schiavon models are younger by several Gyr compared to the 
\citet{ch1tmk04} models.
\label{hgfe}
}
\end{center}}

\vbox{
\begin{center}
\includegraphics[width=\textwidth]{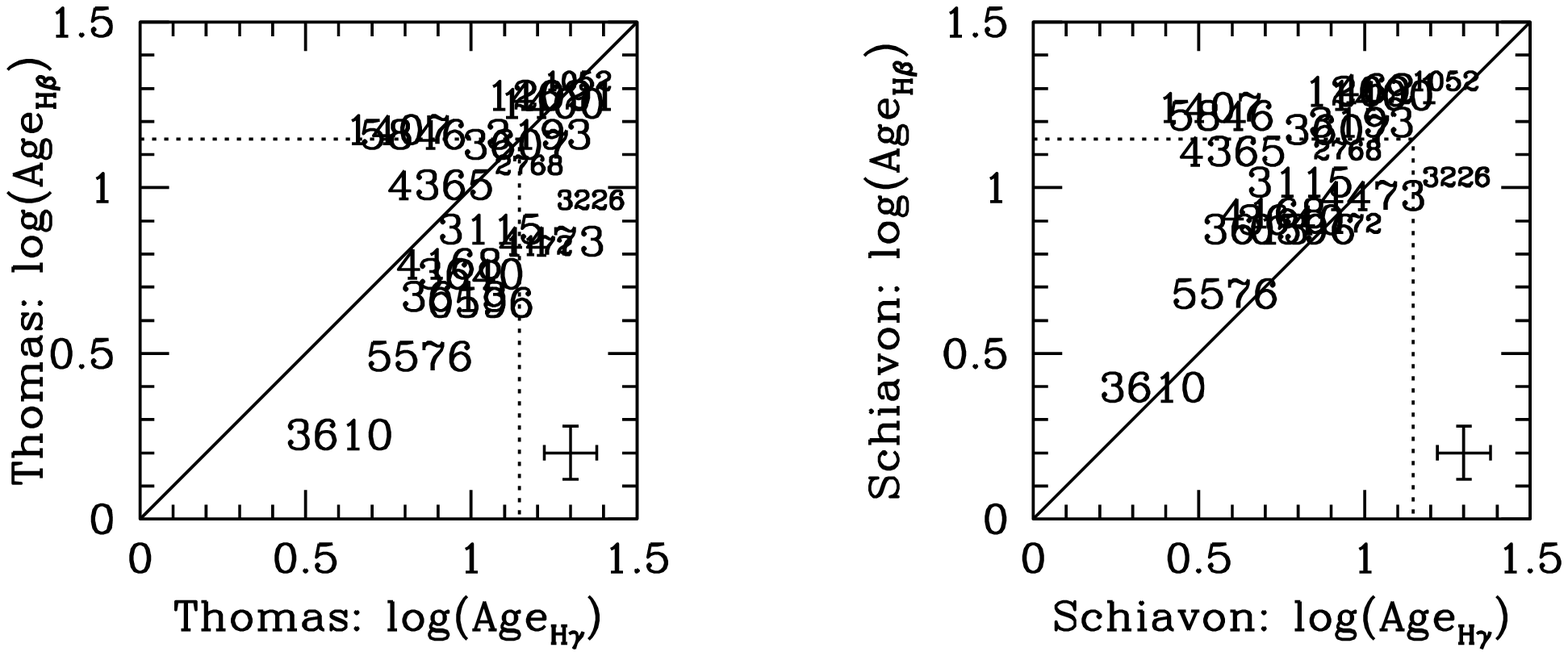}
\figcaption{\small 
Ages derived using the model grids in Figs.~\ref{hbfe} and \ref{hgfe}
are compared.  Measurements in the left panel are based on the Thomas~et~al.
models, while measurements in the right panel are based on the Schiavon models.
As above, galaxies in small type have large emission corrections. 
Typical age uncertainties for the remaining galaxies are presented in
the lower right.  Dotted lines are drawn at Age~$=14$~Gyr on each axis.
The galaxies with inconsistent Balmer measurements
(outliers in Fig.~\ref{balmercomp}) result in discrepant ages using both
sets of models.  The two sets of models are in qualitative agreement,
though as expected from Figs.~\ref{hbfe} and \ref{hgfe} they are in
systematic quantitative disagreement, typically by 2--4~Gyr.  The Thomas~et~al. 
models result in older ages on the H$\beta$ grids (Fig.~\ref{hbfe}) and
younger ages on the H$\gamma_{\rm F}$ grids (Fig.~\ref{hgfe}).  Neither
set of models gives fully consistent ages between the two grids.  The
Schiavon models are more consistant than the Thomas~et~al. models at
young and intermediate ages, while the Thomas~et~al. models are more
consistent for extremely old (Age~$>12$~Gyr) galaxies.
For the majority of galaxies, the difference between ages derived using
different sets of models is significantly greater than the difference 
between ages derived using different Balmer indices.  As a result, the
conclusion is that the choice of model is more important than the choice
of age-sensitive Balmer index in measuring SSP parameters.  
\label{betagamma}
}
\end{center}}

\vbox{
\begin{center}
\includegraphics[width=\textwidth]{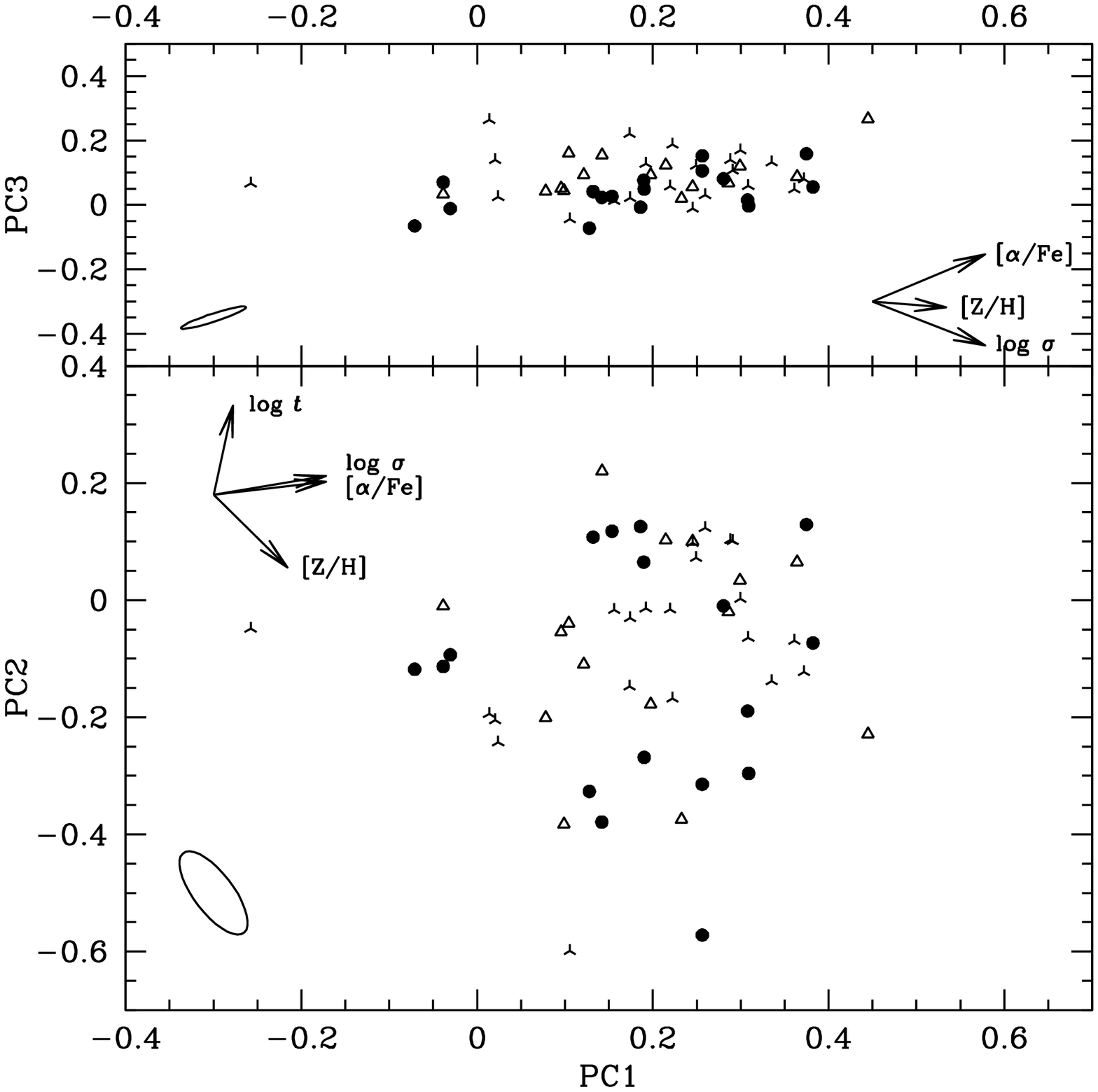}
\figcaption{\small
Principal component projection of the metallicity hyper-plane of
TFWG.  Solid circles are from this study.  The remaining points are
from G93; open triangles are in the volume-limited sample while skeletal
symbols are not.  Error ellipses are shown for the measured index and 
calibration errors for this study (solid ellipses).  The projection
of each SSP parameter in principal component space is shown.  PC1 is primarily
sensitive to log~$\sigma$ and [$\alpha$/Fe], with some metallicity
sensitivity.  PC2 is sensitive to age and metallicity, while PC3 is
depends weakly on both log~$\sigma$ and [$\alpha$/Fe].  Conceptually,
one can think of PC1 measuring a galaxy's position along the 
Mg--$\sigma$ relation, PC2 measuring position along an age--metallicity
relation, and PC3 measuring deviations from the Mg--$\sigma$ relation.
The volume-limited and G93 samples are consistent with being drawn from
the same distributions in each principal component axis.
\label{pc}
}
\end{center}}

\vbox{
\begin{center}
\includegraphics[width=\textwidth]{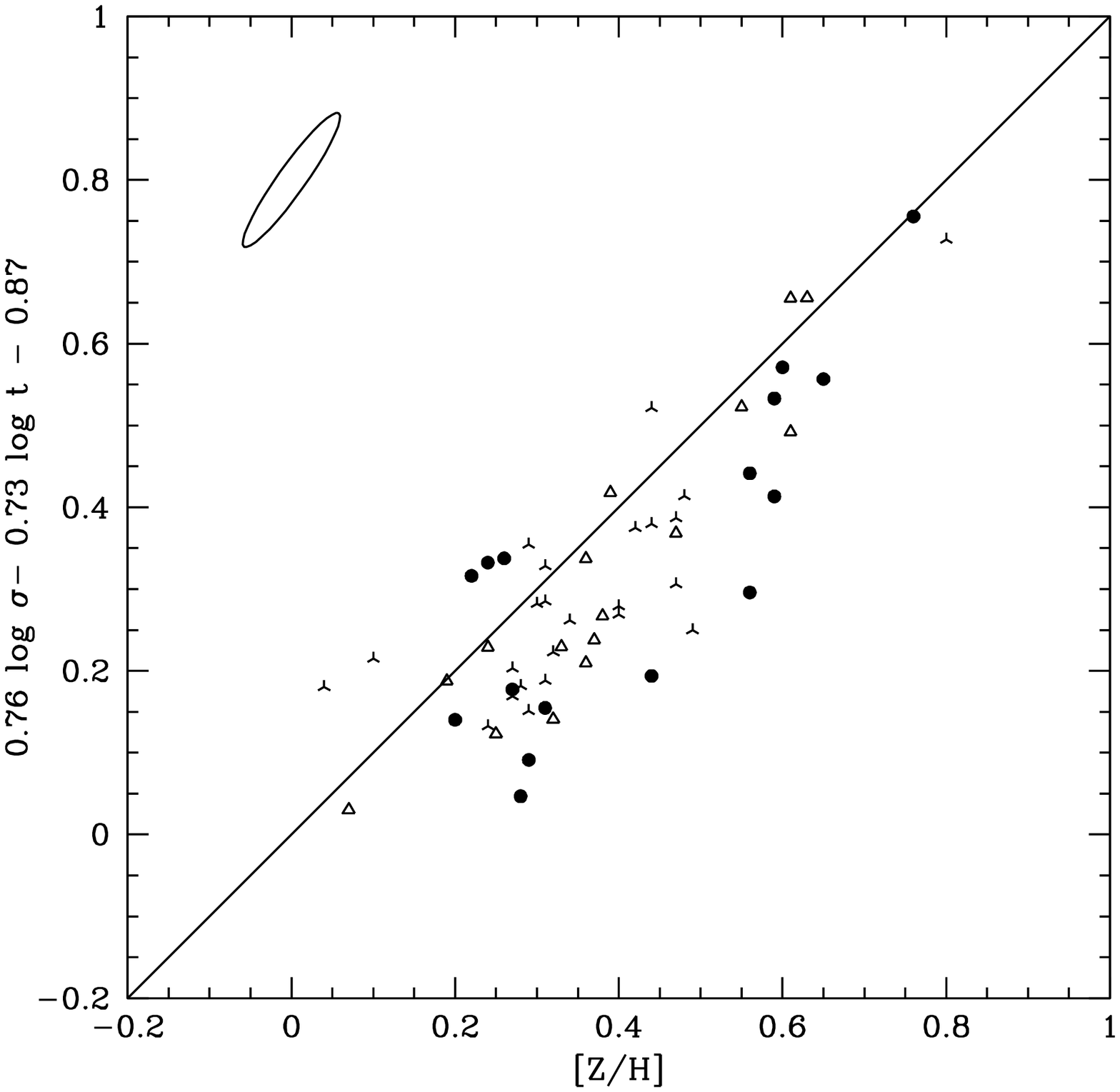}
\figcaption{\small
The $Z$-plane as defined in TFWG (solid line).  Points are as in Fig~\ref{pc}.  
A typical $1\sigma$ error ellipse is shown in the upper left.  The 
normalization of this relation between velocity dispersion, age, and
metallicity depends on the models used, as would be expected based on 
Fig.~\ref{diff}.  The difference in models readily explains the offset
between the TFWG fit and the data.  The most metal-rich galaxy is NGC~3610,
which is examined in detail in \citet{3610}.  The volume-limited and G93
samples are consistent with being drawn from the same distribution along
this edge-on projection of the $Z$-plane.
\label{zplane}
}
\end{center}}

\vbox{
\begin{center}
\includegraphics[angle=-90,width=\textwidth]{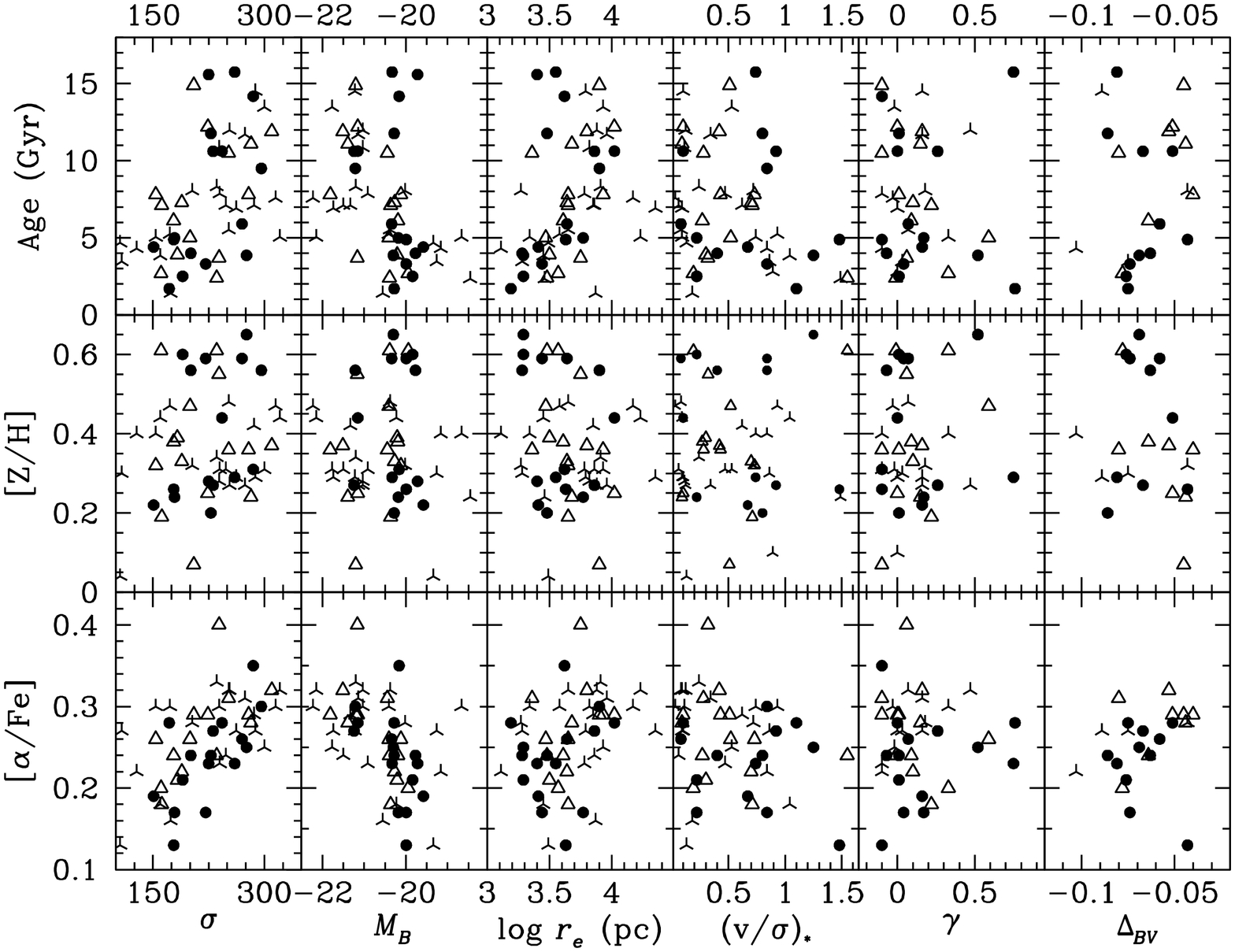}
\figcaption{\small
Comparison between parameters relating to the average stellar composition
(age, [Z/H], [$\alpha$/Fe]) and parameters relating to galaxy size ($\sigma$,
$M_B$, log~$r_e$), shape ($(v/\sigma)_*$), density profile (the power-law 
slope $\gamma$), and color gradient.  Points are as in Fig~\ref{pc}.  
Several well-known trends
are apparent.  The [$\alpha$/Fe]--$\sigma$ relation is directly analogous to
the Mg--$\sigma$ relation (Fig.~\ref{index}).  It is also clear that the
faintest and least massive galaxies are all relatively 
young, while brighter and more massive galaxies span a wide range of ages.
\label{asig}}
\end{center}}

\vbox{
\begin{center}
\includegraphics[width=\textwidth]{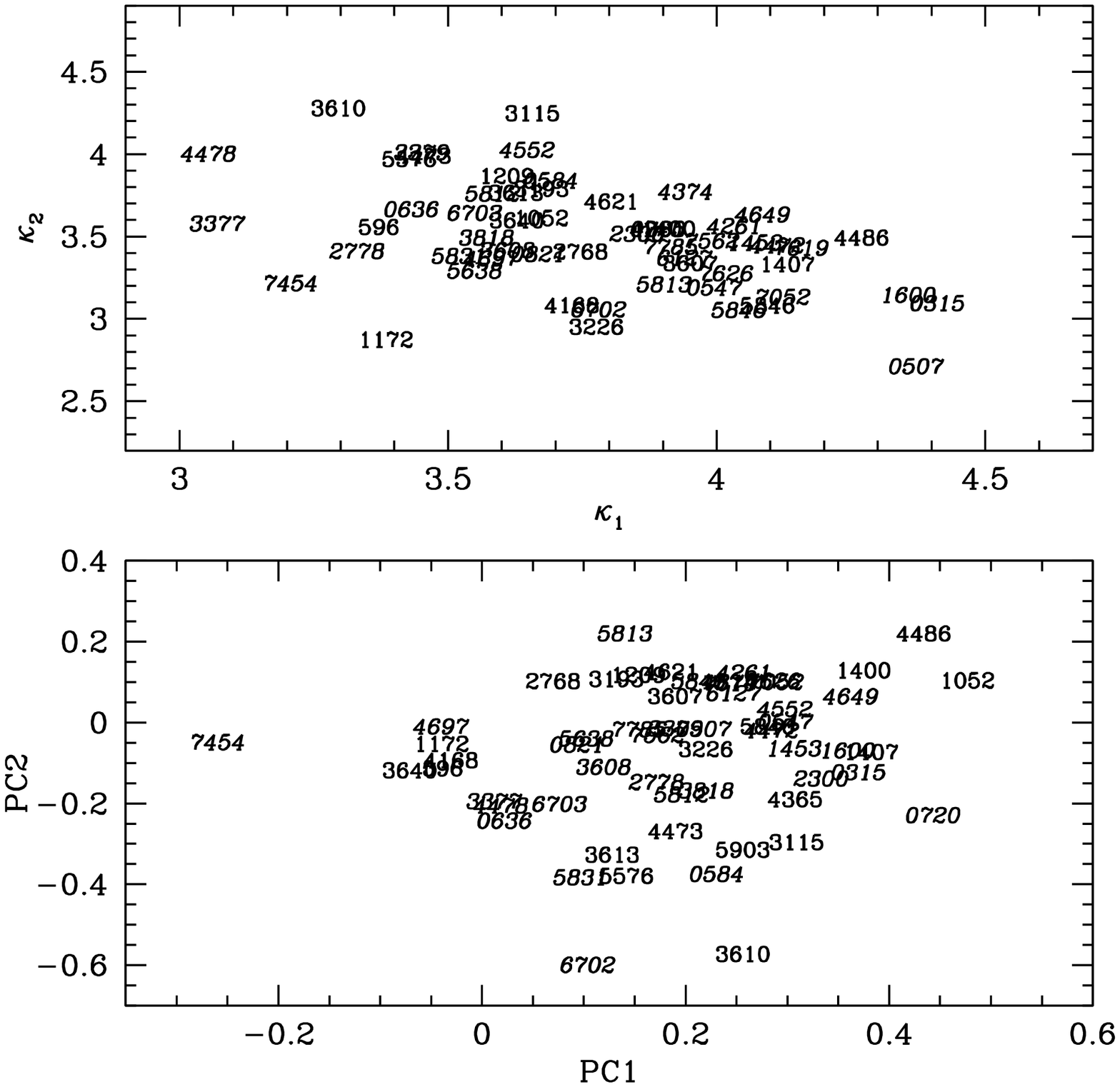}
\figcaption{\small
The face-on projections of the Fundamental Plane (top panel) and $Z$-plane
(bottom panel) are shown.  Galaxies in normal type are from this work, while
those from G93 are in italic type.  No mapping from
one plane to the other is found.  
\label{kappc}
}
\end{center}}


\clearpage

\begin{thebibliography}{DUM}
%
\bibitem[Baade \& Gaposchkin(1963)]{ch1baade} Baade, W.~\& 
Gaposchkin, C.~H.~P.\ 1963, Cambridge, Harvard University Press, 1963.,
%
\bibitem[Barnes \& Hernquist(1996)]{bh96} Barnes, J.~E.~\& 
Hernquist, L.\ 1996, \apj, 471, 115 
%
\bibitem[Bassin \& Bonatto(2003)]{bassin} Bassin, M., \& 
Bonatto, C.\ 2003, \aap, 410, 803
%
\bibitem[Bender et~al.(1992)Bender, Burstein, \& Faber]{kappa} Bender, R., 
Burstein, D., \& Faber, S.~M.\ 1992, \apj, 399, 462
%
\bibitem[Bender et~al.(1993)Bender, Burstein, \& Faber]{mgsig} Bender, R., 
Burstein, D., \& Faber, S.~M.\ 1993, \apj, 411, 153
%
\bibitem[Beuing et~al.(2002)]{beuing02} Beuing, J., Bender, R., Mendes de 
Oliveira, C., Thomas, D., \& Maraston, C. 2002, \aap, 395, 431
%
\bibitem[Brodie et al.(2005)]{brodie05} Brodie, J.~P., Strader, 
J., Denicol{\' o}, G., Beasley, M.~A., Cenarro, A.~J., Larsen, S.~S., 
Kuntschner, H., \& Forbes, D.~A.\ 2005, \aj, 129, 2643
%
\bibitem[Burkert(1993)]{burkert} Burkert, A.\ 1993, \aap, 278, 23
%
\bibitem[Carollo et~al.(1997)]{carollo97} 
Carollo, C.~M., Danziger, I.~J., Rich, R.~M., \& Chen, X.\ 1997, \apj, 491, 
545
%
\bibitem[Caldwell et al.(2003)]{crc03} Caldwell, N., Rose, 
J.~A., \& Concannon, K.~D.\ 2003, \aj, 125, 2891
%
\bibitem[Cowie et~al.(1996)]{cowie} Cowie, L.~L., Songaila, A., Hu, E.~M., 
\& Cohen, J.~G.\ 1996, \aj, 112, 839
%
\bibitem[Denicol{\' o} et al.(2005)]{ch1denicolo} Denicol{\' o}, 
G., Terlevich, R., Terlevich, E., Forbes, D.~A., Terlevich, A., \& 
Carrasco, L.\ 2005, \mnras, 356, 1440
%
\bibitem[de Vaucouleurs(1948)]{dv48} de Vaucouleurs, G. 1948, Ann. Astrophys.,
11, 247
%
\bibitem[Faber et~al.(1989)]{ch1faber89} Faber, S.~M., Wegner, G., Burstein, D.,
Davies, R.~L., Dressler, A., Lynden-Bell, D., \& Terlevich, R.~J. 1989,
\apjs, 69, 763
%
\bibitem[Faber et al.(1997)]{faber97} Faber, S.~M., et al.\ 
1997, \aj, 114, 1771
%
\bibitem[Gebhardt et al.(2003)]{geb03} Gebhardt, K., et al.\ 
2003, \apj, 597, 239
%
\bibitem[Gonzalez(1993)]{ch1g93} Gonzalez, J.~J. 1993, Ph.D. thesis, University
of California, Santa Cruz (G93)
%
\bibitem[Howell(2005)]{paper2} Howell, J.~H. 2005, submitted
%
\bibitem[Howell et al.(2004)]{3610} Howell, J.~H., Brodie,
J.~P., Strader, J., Forbes, D.~A., \& Proctor, R.\ 2004, \aj, 128, 2749
%
\bibitem[Idiart et~al.(2002)Idiart, Michard \& de Freitas Pacheco]{ch1idiart02}
Idiart, T.~P., Michard, R. \& de Freitas Pacheco, J.~A.\ 2002, \aap, 383,
30
%
\bibitem[Jones(1999)]{j99} Jones, L.A. 1999, PhD Thesis, University of
North Carolina
%
\bibitem[J{\o}rgensen(1997)]{ch1jorg97} J{\o}rgensen, I.\ 1997, \mnras,
288, 161
%
\bibitem[J{\o}rgensen(1999)]{jorg99} J{\o}rgensen, I.\ 1999, 
\mnras, 306, 607
%
\bibitem[Kaiser(1960)]{kaiser} Kaiser, H.~F. 1960, Education and Psychological
Measurement, 20, 141
%
\bibitem[Kuntschner \& Davies(1998)]{ch1kd98} Kuntschner, H.~\&
Davies, R.~L.\ 1998, \mnras, 295, L29
%
\bibitem[Kuntschner et al.(2002)]{ch1kuntschner} Kuntschner, H.,
Smith, R.~J., Colless, M., Davies, R.~L., Kaldare, R. \& Vazdekis, A.\
2002, \mnras, 337, 172
%
\bibitem[Kuntschner et al.(2002)]{3115ref} Kuntschner, H.,
Ziegler, B.~L., Sharples, R.~M., Worthey, G., \& Fricke, K.~J.\ 2002, \aap,
395, 761
%
\bibitem[Larsen et al.(2003)]{4365ref} Larsen, S.~S., Brodie,
J.~P., Beasley, M.~A., Forbes, D.~A., Kissler-Patig, M., Kuntschner, H., \&
Puzia, T.~H.\ 2003, \apj, 585, 767
%
\bibitem[Larson(1974)]{ch1larson} Larson, R.~B.\ 1974, \mnras, 
166, 585
%
\bibitem[Lauer et~al.(2005)]{lauerprep} Lauer, T.~R., et~al. 2005, \aj, 
129, 2138
%
\bibitem[Longhetti et~al.(2000)]{longhetti} Longhetti, M., Bressan, A., Chiosi, 
C., \& Rampazzo, R.\ 2000, \aap, 353, 917
%
\bibitem[Puzia et al.(2002)]{4365ir} Puzia, T.~H., Zepf,
S.~E., Kissler-Patig, M., Hilker, M., Minniti, D., \& Goudfrooij, P.\ 2002,
\aap, 391, 453
%
\bibitem[Puzia(2003)]{ch1puzia} Puzia, T.~H.\ 2003, Ph.D.~Thesis, 
Ludwig-Maximilians-Universitaet Muenchen
%
\bibitem[Puzia et al.(2004)]{puzia} Puzia, T.~H., Kissler-Patig, M., Thomas,
D., Maraston, C., Saglia, R. P., Bender, R., Richtler, T., Goudfrooij, P.,
\& Hempel, M.  2004, \aap, 415, 123
%
\bibitem[Rich(1998)]{ch1bwid} Rich, R.~M. 1998, ASP Conf.\ Ser.\ 147, Abundance
Profiles: Diagnostic Tools for Galaxy History, ed. D.~Friedli, M.~Edmunds,
C.~Robert, \& L.~Drissen (San Francisco: ASP), p.~36
%
\bibitem[Rose et al.(1994)]{rose} Rose, J.~A., Bower, R.~G., 
Caldwell, N., Ellis, R.~S., Sharples, R.~M., \& Teague, P.\ 1994, \aj, 108, 
2054
%
\bibitem[Schechter(1976)]{schechter} Schechter, P.\ 1976, \apj, 
203, 297
%
\bibitem[Schiavon (2005)]{ch1schiavon} Schiavon, R. 2005, in preparation
%
\bibitem[Spaans \& Carollo(1997)]{sc97} Spaans, M.~\& 
Carollo, C.~M.\ 1997, \apjl, 482, L93
%
\bibitem[Strader \& Brodie (2004)]{ch1strader04} Strader, J. \& Brodie, J. 2004,
\aj, 128, 1671
%
\bibitem[Tantalo, Chiosi, \& Bressan(1998)]{tantalo} Tantalo, 
R., Chiosi, C., \& Bressan, A.\ 1998, \aap, 333, 419
%
\bibitem[Terlevich \& Forbes(2002)]{tf02} Terlevich, A.~I., 
\& Forbes, D.~A.\ 2002, \mnras, 330, 547
%
\bibitem[Thomas et~al.(2003)Thomas, Maraston, \& Bender]{ch1tmb03}
Thomas, D., Maraston, C., \& Bender, R.  2003, \mnras, 339, 897
%
\bibitem[Thomas et~al.(2004)Thomas, Maraston, \& Korn]{ch1tmk04} Thomas, D., 
Maraston, C., \& Korn, A.\ 2004, \mnras, 351, L19
%
\bibitem[Tonry et~al.(2001)]{ch1tonry01} Tonry, J.~L., Dressler, A., Blakeslee, 
J.~P., Ajhar, E.~A., Fletcher, A.~B., Luppino, G.~A., Metzger, M.~R., \&
Moore, C.~B. 2001, \apj, 546, 681
%
\bibitem[Trager et~al.(1998)]{ch1trager98} Trager, S.~C., Worthey, G., Faber,
S.~M., \& Gonzalez, J.~J. 1998, \apjs, 116, 1
%
\bibitem[Trager et~al.(2000a)]{ch1tfwg1} Trager, S.~C., Faber, S.~M.,
Worthey, G., \& Gonzalez, J.~J. 2000a, \aj, 119, 1645
%
\bibitem[Trager et~al.(2000b)]{ch1tfwg2} Trager, S.~C., Faber, S.~M.,
Worthey, G., \& Gonzalez, J.~J. 2000b, \aj, 120, 165 (TFWG)
%
\bibitem[Tripicco \& Bell(1995)]{ch1tb95} Tripicco, M.~J.~\& 
Bell, R.~A.\ 1995, \aj, 110, 3035 
%
\bibitem[van der Marel(1994)]{ch1pixfit} van der Marel 1994, MNRAS, 270, 271
%
\bibitem[Worthey et~al.(1994)]{ch1worthey94} 
Worthey, G., Faber, S.~M., Gonzalez, J.~J., \& Burstein, D.\ 1994, \apjs, 
94, 687
%
\bibitem[Worthey(1994)]{w94model} Worthey, G.\ 1994, \apjs, 95, 107
%
\bibitem[Worthey \& Ottaviani(1997)]{ch1wo97} Worthey, G., \& Ottaviani, D.~L.
1997, \apjs, 111, 377
%
\bibitem[Zepf \& Ashman(1993)]{zepf93} Zepf, S.~E.~\& Ashman, K.~M.\ 1993,
\mnras, 264, 611
%
\end{thebibliography}
\end{document}